\documentclass{article}
\usepackage{latexsym,amssymb}
\usepackage{graphics}
\usepackage{pstricks}
\usepackage{amsmath}
\usepackage{subfigure}
\usepackage{vmargin}  
\usepackage{epsfig}

\setmarginsrb{2.cm}{2.cm}{2.cm}{2.5cm}{0cm}{0.cm}{0cm}{1cm}

\begin{document}

\begin{flushright}
PCCF-RI-0204\\
CPPM-P-2002-01\\
July 2002
\end{flushright}

\vspace{1.cm}

\begin{center}
{\Large {\bf Neutrino Indirect Detection of Neutralino Dark Matter in the}}
\vspace{0.2cm}\\
{\Large {\bf CMSSM.}}
\vspace{0.3cm}\\
{\large V. Bertin $^1$, E. Nezri $^{1 \ 2}$, J. Orloff $^2$}
\vspace{0.3cm}\\
$^1$ Centre de Physique des Particules de Marseille\\
IN2P3-CNRS, Universit\'e de la M\'editerran\'ee, F-13288 Marseille Cedex 09
\vspace{0.2cm}\\
$^2$ Laboratoire de Physique Corpusculaire de Clermont-Ferrand\\ 
IN2P3-CNRS, Universit\'e Blaise Pascal, F-63177 Aubiere Cedex
\vspace{0.2cm}\\
{\tt email : bertin@cppm.in2p3.fr, nezri@in2p3.fr, orloff@in2p3.fr}

\end{center}

\abstract{We study potential signals of neutralino dark matter indirect
    detection by neutrino telescopes in a wide range of CMSSM parameters. We
    also compare with direct detection potential signals taking into account
    in both cases present and future experiment sensitivities. Only models
    with neutralino annihilation into gauge bosons can satisfy 
    cosmological constraints and current neutrino indirect detection sensitivities. For both direct and indirect detection, only next generation experiments will be able to really test this kind of models.}


\section{Introduction}
Our present understanding of the universe is described in the framework of
general relativity and cosmology. The densities of its components are
related by \cite{Peacock}:
\begin{equation}
\Omega(a)-1=\frac{\Omega_{tot}-1}{1-\Omega_{tot}+\Omega_{\Lambda}
a^{2}+\Omega_{mat} a^{-1}+\Omega_{rel}a^{-2}} \ ,
\end{equation}
where $a$ is the scale factor.  This equation and current experimental
results suggest and focus on a flat universe with the following parameters
\cite{Lineweaver:2001}:\\
\begin{center}
\begin{tabular}{lll}
\multicolumn{2}{l}{Cosmological constant:} &$\Omega_{\Lambda}=0.7 \pm 0.1 $ \\
Matter:& &$\Omega_{mat}=0.3 \pm 0.1 $ \\
  &baryonic matter: &$\Omega_{b}=0.04 \pm 0.01 $; 
	including $\Omega_{vis}\lesssim0.01$ \\
  &cold dark matter: & $\Omega_{DM}=0.26 \pm 0.1 $ \\
\multicolumn{2}{l}{Relativistic components:} & $0.01\lesssim
	\Omega_{rel}\lesssim0.05$\\
  &neutrinos: &$0.01\lesssim\Omega_{\nu}\lesssim0.05 $\\
  &photons:   &$\Omega_{\gamma}=4.8^{+1.3}_{-0.9} \times10^{-5} $\\
\multicolumn{2}{l}{Hubble's constant:} &$ h \equiv H_0/100 
	\ {\rm km^{-1}\ s^{-1}\ Mpc^{-1}}=0.72\pm0.08$ \\
\end{tabular}
\end{center}

In the framework of the Minimal Supersymmetric Standard Model (MSSM)
\cite{Fayet:1977cr,Barbieri:1988xf,Martin:1997ns,Bagger:1996ka,Haber:1985rc},
the lightest supersymmetric particle (LSP) is typically the lightest
of the neutralinos $\chi_1(\equiv\chi),\chi_2,\chi_3,\chi_4 $, the
mass eigenstates of the neutral gauge and Higgs boson superpartners ($
\tilde{B},\tilde{W^3},\tilde{H^0_d}, \tilde{H^0_u} $). In the rest of
this paper, the LSP is thus assumed to be the lightest neutralino
$\chi_1$ and is called more generically $\chi$. Assuming R-parity
conservation ($ R\equiv(-1)^{B+L+2S} $), the neutralino is a good
stable candidate for cold dark matter \cite{Jungman:1996df}. In this
context, all sparticles produced after the big-bang give a neutralino
in their decay chain, leading to a relic bath of $\chi$ in the present
universe. These neutralinos could be observed via direct detection
($\chi$ interaction with a nucleus of a detector), or indirect
detection of their annihilation products ($\nu$, $\gamma$, $\bar{p}$,
$\bar{D}$, $e^+$). We will mainly focus in this paper on $\nu$
indirect detection. The relic neutralinos are gravitationally captured
by massive astrophysical bodies and accumulated at the centre of these
objects by successive elastic scatterings on their nuclei. The
captured neutralino population annihilates and gives rise to neutrino
fluxes which could be detectable in neutrino telescopes like
Amanda/Icecube, Antares, Baikal.\\

\section{Neutralino in the CMSSM}

In the MSSM, the mass matrix of neutralinos in the basis $(
-i\tilde{B},-i\tilde{W^3},\tilde{H^0_d}, \tilde{H^0_u} )$ is
\begin{equation}
M_{\chi,\chi_{2},\chi_3,\chi_4}=\left(\begin{array}{cccc}
M_1 & 0 & -m_Z \cos\beta \sin\theta_W & m_Z \sin\beta \sin\theta_W \\
0& M_2 & m_Z \cos\beta \cos\theta_W & -m_Z \sin\beta \cos\theta_W \\
-m_Z \cos\beta \sin\theta_W & m_Z \cos\beta \cos\theta_W & 0 &-\mu\\
m_Z \sin\beta \sin\theta_W & -m_Z \sin\beta \cos\theta_W &-\mu&0\\
\end{array}
\right)
\end{equation}
where $M_1$, $M_2$ are mass term of $U(1)$ and $SU(2)$ gaugino fields,
$\mu$ is the higgsino ``mass'' parameter and $\tan{\beta}=<H_u>/<H_d>$ is
the ratio of the neutral Higgs vacuum expection values. The neutralino
composition is:
\begin{equation}\chi=N_1 \tilde{b}+N_2\tilde{W^3}+N_3\tilde{H^0_d}+N_4
  \tilde{H^0_u} 
\label{neutralino} 
\end{equation}
and we define its gaugino fraction as $g_{frac}=|N_1|^2+|N_2|^2$.

In this model, the introduction of soft terms in the Lagrangian breaks
explicitly supersymmetry, leading to a low energy effective theory with 106
parameters. The MSSM is therefore not very predictive, and a non biased
exploration of its parameter space is not possible. As a first step, we
will therefore as usual concentrate on gravity-mediated supersymmetry
breaking in supergravity \cite{Nilles:1984ge} inspired models, with Grand
Unification of soft terms at $E_{GUT}\sim2.10^{16}$ GeV parameterized by
$m_0$ (common scalar mass), $m_{1/2}$ (common gaugino mass) and $A_0$
(common trilinear term). Together with $\tan\beta$ and $sgn(\mu)$, these
define a 5 parameters constraint MSSM (CMSSM) or mSugra model
\cite{Chamseddine:1982jx,Barbieri:1982eh,Hall:1983iz}, from which the 106
parameters can be deduced through renormalization group equations (RGE).

Due to the large top Yukawa coupling, renormalization group evolution can
drive $m^2_{H_u}\vert_{Q_{EWSB}}$ and/or $ m^2_{H_d}\vert_{Q_{EWSB}}$ to
negative values, so that the electroweak symmetry breaking (EWSB)
originates purely in quantum corrections, which realizes radiative
electroweak symmetry breaking. Minimization of the scalar potential at the
electroweak breaking scale $Q_{EWSB}$ yields the condition:
\begin{equation}\label{potminimi}
\frac{1}{2}m^2_Z=\frac{m^2_{H_d}\vert_{Q_{EWSB}}
	-m^2_{H_u}\vert_{Q_{EWSB}}\tan^2\beta}
	{\tan^2\beta-1}-\mu^2\vert_{Q_{EWSB}}
	{\underset{\tan{\beta}\gtrsim5}\sim}
	-m^2_{H_u}\vert_{Q_{EWSB}}-\mu^2\vert_{Q_{EWSB}}
	\ ,
\end{equation}
where usually $ Q_{EWSB}$ is taken as $\sqrt{m_{\tilde{t}_1}
m_{\tilde{t}_2}}$ \cite{Gamberini:1990jw} to minimize one loop corrections.

Such mSugra models offer the theoretical advantage over generic MSSM models
that problems such as Landau poles, charge and color breaking (CCB) minima
are partially addressed when dealing with RGE.

\begin{figure}[!ht]
 \begin{center} \begin{tabular}{c}
 \includegraphics[width=\textwidth]{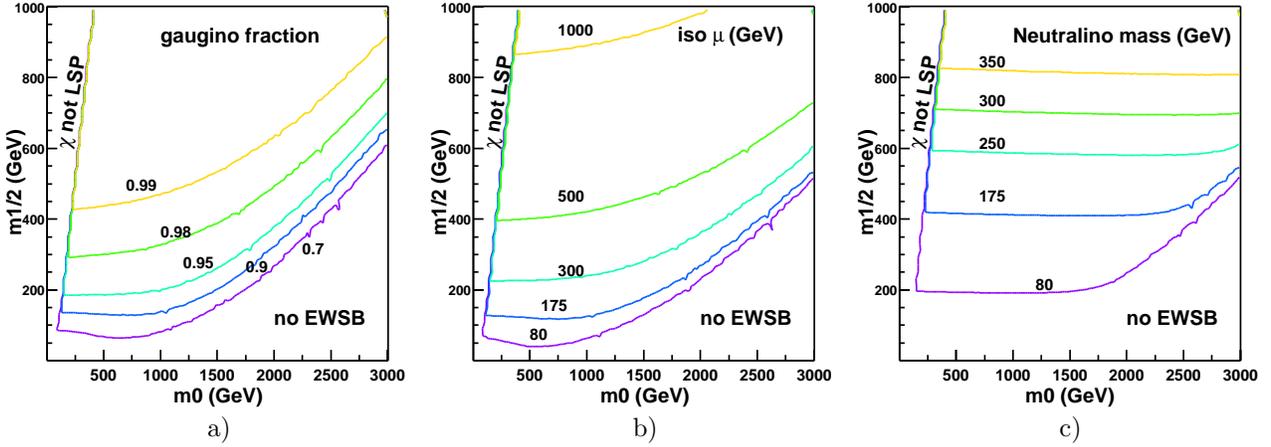}\\ a)
 \hspace{0.3\textwidth} b) \hspace{0.3\textwidth} c) \end{tabular}
 \caption{\small a) Gaugino fraction, b) value of $\mu$ and c) neutralino
 mass in the $(m_0,m_{1/2})$ plane.  Region where $\chi$ is not the LSP or
 EWSB is not achieved are also indicated.}  \label{isomass} \end{center}
\end{figure}

In mSugra, the lightest neutralino can exhibit two different natures,
depending on the input parameter values \cite{Feng:1999zg,Feng:2000gh}:
\begin{itemize}
\item[-] an almost pure bino-like neutralino for low $m_0$, as the RGE
drive $M_1\vert_{Q_{EWSB}} \simeq 0.41M_1\vert_{GUT} = 0.41m_{1/2} <<
|\mu|_{Q_{EWSB}}$ and $M_2\vert_{Q_{EWSB}} \simeq 0.83M_1\vert_{GUT} =
0.83m_{1/2} << |\mu|_{Q_{EWSB}}$;
\item[-] for $m_0\gtrsim1000$ GeV, the neutralino picks up some higgsino
mixing for $\tan{\beta}\gtrsim5$ as the increase of $m_0$ drives
$m^2_{H_u}$ to less negative values so that both $|m^2_{H_u}|$ and $|\mu|$
(via eq. \ref{potminimi}) decrease. One can then have $ |\mu|_{Q_{EWSB}}
\lesssim M_1\vert_{Q_{EWSB}} $ depending on $m_{1/2}$. When $|\mu|$ is too
small, EWSB cannot be achieved.
\end{itemize}
In addition, the very low $m_0$ values can lead to models with tachyonic
sfermions (low $m_{1/2}$) or a slepton LSP. These behaviors are shown on
figure \ref{isomass} in the $(m_0,m_{1/2})$ plane. The neutralino mass is
$\sim 0.4m_{1/2}$ but is affected by low $\mu$ values at high $m_0$. Unless
explicitly quoted we use $m_t=174.3$ GeV.

In this paper, the RGE, SUSY spectrum and potential minimization are derived
using the SUSPECT program \cite{Suspect,Djouadi:2001yk} version 2.005 which
includes 2 loops RGE \cite{Castano:1994ri}, the tadpole method to minimize the
scalar potential \cite{Barger:1994gh} and all radiative corrections in the
mass spectrum \cite{Pierce:1997zz}. The neutralino relic density, including
$\chi\chi^+$ and $\chi\chi_2$ coannihilations, and detection signals are
calculated by the DarkSUSY (DS) package \cite{Gondolo:2000ee,Darksusy}
\footnote{These calculations use a neutralino local density
  $\rho_{\chi}=0.3 \ {\rm GeV/cm^3}$ and a Maxwellian velocity distribution
  with $v_0=220 \ {\rm km \ s^{-1}}$}
which correctly implements $s$-channel poles and thresholds
\cite{Nihei:2001qs}. Our interface (available from nezri@in2p3.fr,
submitted for inclusion in DS) shortcuts the DS spectrum recalculation, to
comply with the SUSPECT spectra which have been checked
\cite{Allanach:2001hm} against SOFTSUSY \cite{Allanach:2001kg}.

\section{Branching ratios in neutralino annihilation}
\begin{figure}[hbtp]
\begin{center}
 \begin{tabular}{ccccc}
\includegraphics[width=2.cm]{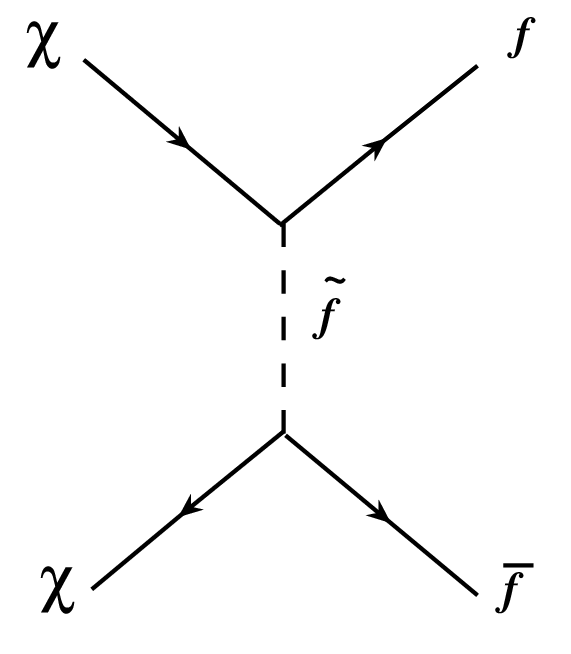}&
\includegraphics[height=2.cm]{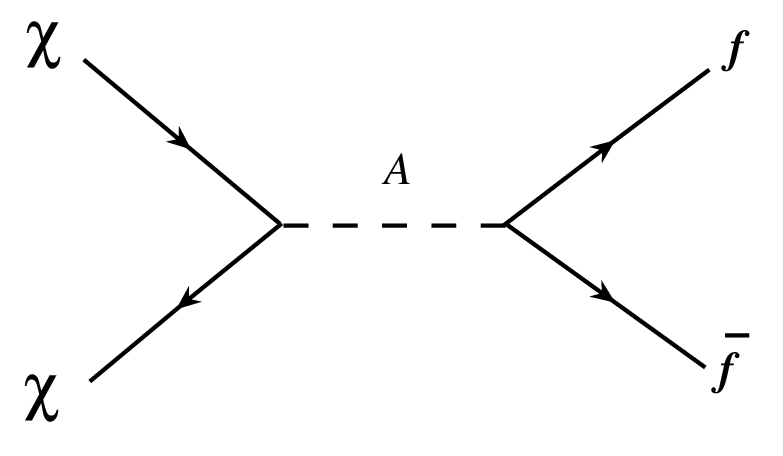}&
\includegraphics[height=2.cm]{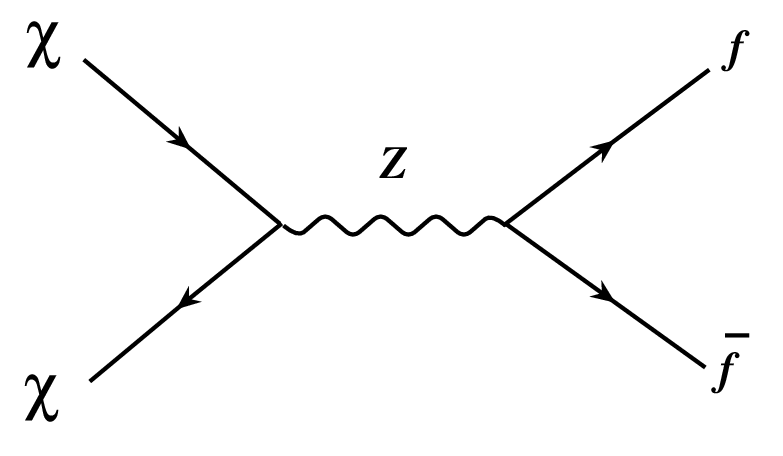}&
\includegraphics[width=2.cm]{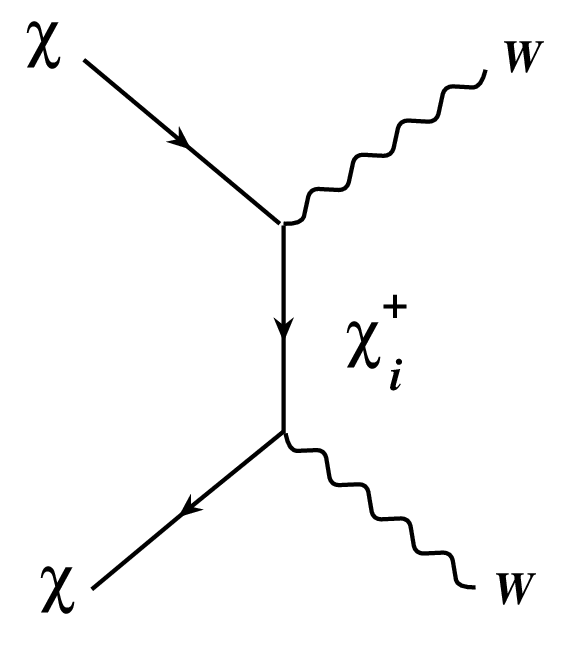}&
\includegraphics[width=2.cm]{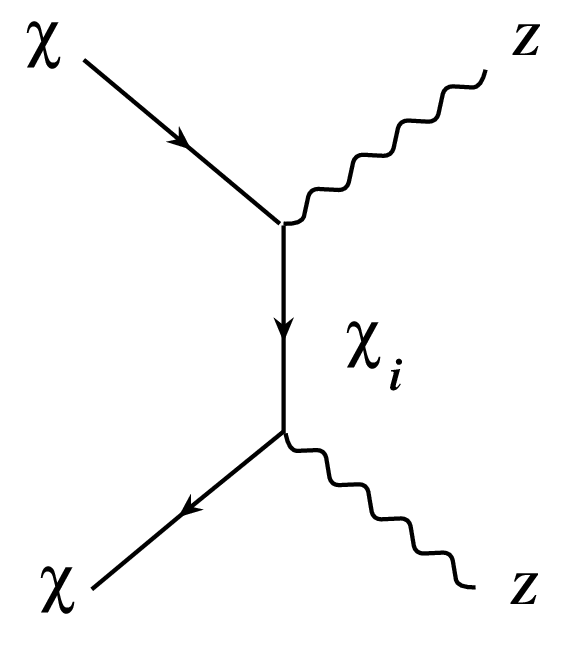}\\
 $A^{\tilde{f}}_{f\bar{f}}\varpropto(\frac{m_{\chi}}{m_{\tilde{f}}})^2$ &
 $A^A_{f\bar{f}} \varpropto \frac{\tan{\beta}m_{f_{d}}}{m_{W}} 
	(\frac{m_{\chi}}{m_A})^2 N_1N_{3(4)}$ &
 $A^Z_{f\bar{f}}\varpropto(\frac{m_fm_{\chi}}{m^2_Z})N^2_{3(4)}$ &
\multicolumn{2}{c}{$A^{\chi_i/\chi^{+}_i}_{W^{+}W^{-}/ZZ} \varpropto 
	\frac{1}{1+(m_{\chi^{+}_i}/m_{\chi})^2-(m_{W}/m_{\chi})^2}$}\\
&
$\tan{\beta}m_{f_{d}}\leftrightarrow \frac{m_{f_{u}}}{\tan{\beta}}$ &
&
\multicolumn{2}{c}{$(m_{\chi^{+}_i},m_W)\leftrightarrow (m_{\chi_i},m_Z)$}
\end{tabular}
 \caption{\small Leading channels in neutralino annihilation at rest, with
   the parametric dependance of their amplitude, $N_i$ being defined in
   equation \ref{neutralino} \cite{Jungman:1996df}.}  \label{channels}
   \end{center}
\end{figure}

\begin{figure}[!ht]
 \begin{center}
 \includegraphics[width=0.5\textwidth]{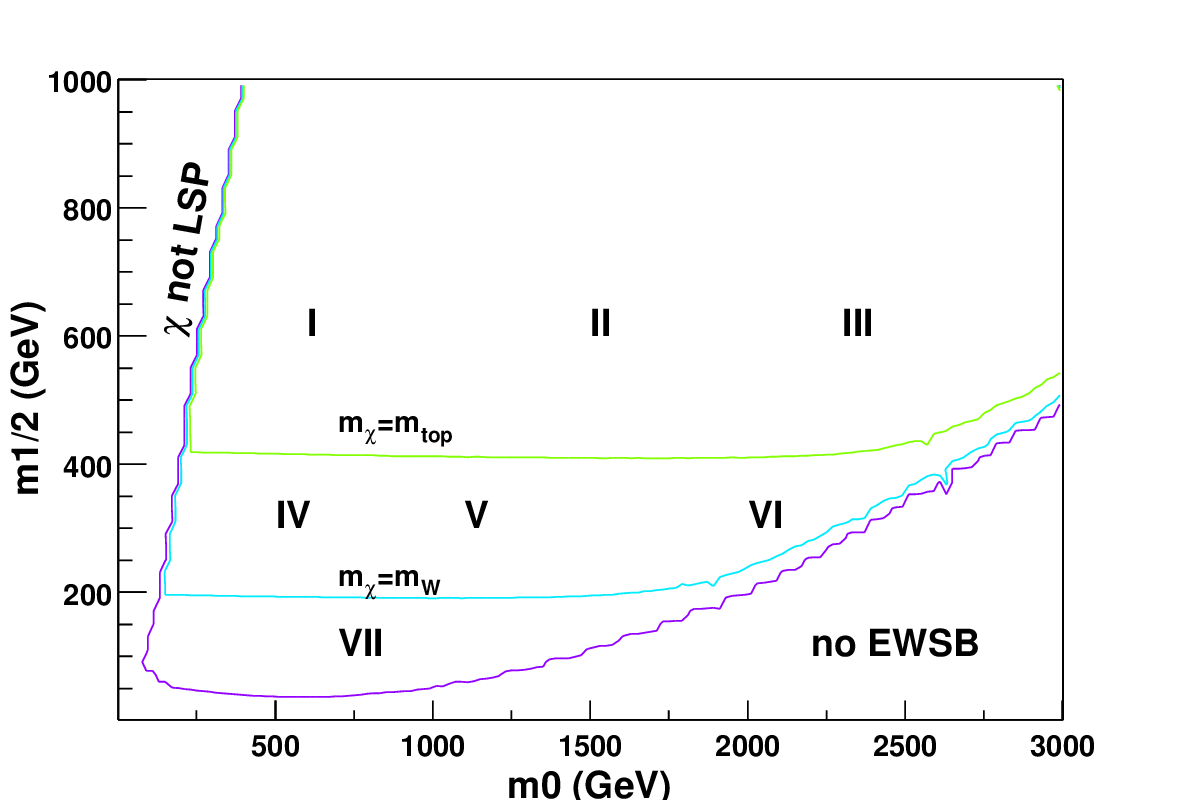}
 \caption{\small Typical regions of neutralino annihilation.}
 \label{schemam0mhalf}
 \end{center}
\end{figure}

The leading processes in neutralino annihilation at rest are shown in
figure \ref{channels}, where the last two processes are proportional to the
higgsino/wino fraction of $\chi$ and $\chi^{(+)}_i$. In mSugra models, one
can typically distinguish different regions of annihilation branching
ratios in the $(m_0,m_{1/2}) $ plane, for almost any values of $
\tan{\beta} (\geq 5)$ (see figure \ref{schemam0mhalf}).

{\bf Above top thresholds:} \emph{i.e.} $m_{\chi} \gtrsim m_{top} \Leftrightarrow
m_{1/2}\gtrsim400$ GeV or slightly higher in the mixed higgsino region.

\begin{figure}[!ht]
 \begin{center} \includegraphics[width=\textwidth]{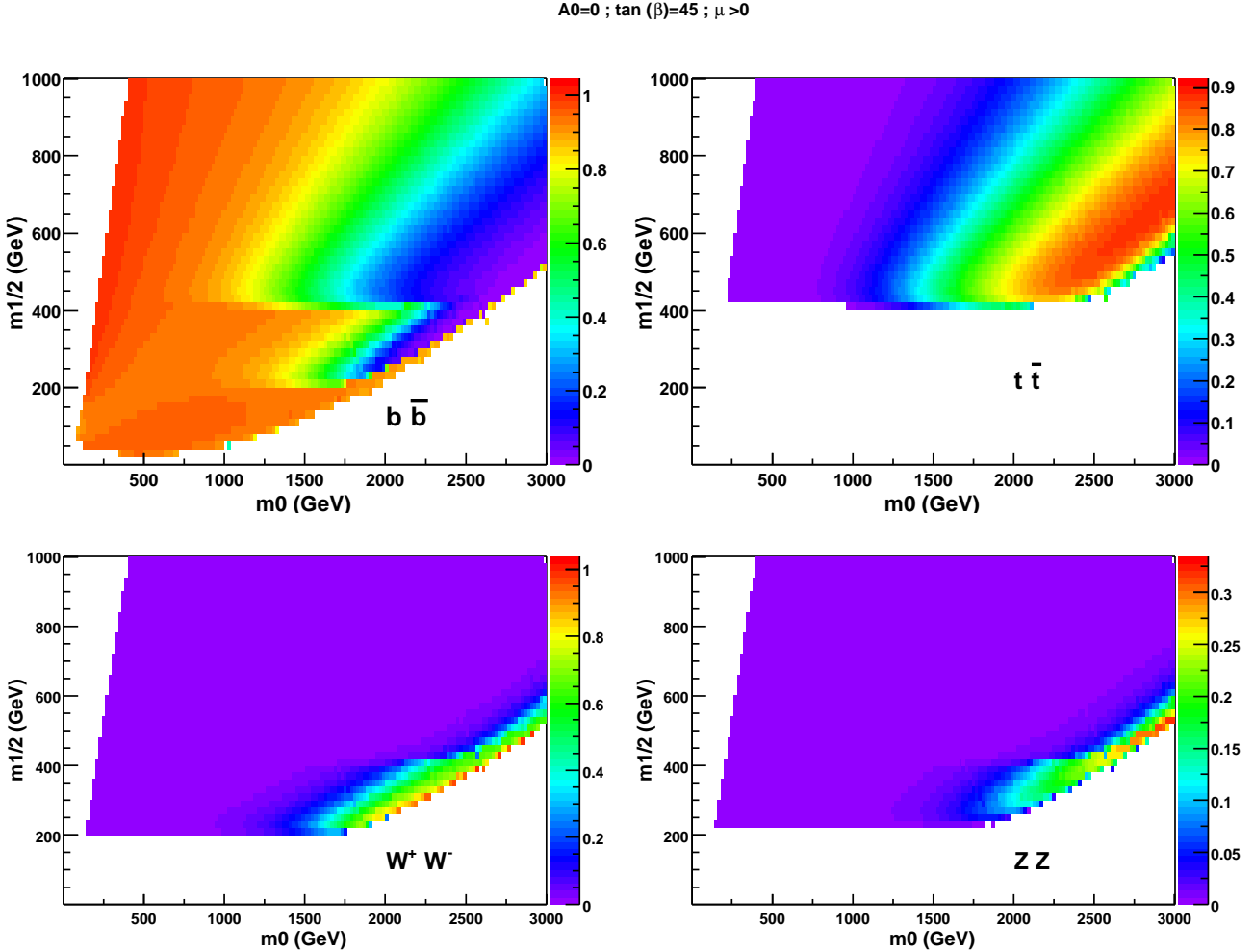}
 \caption{\small Dominant branching ratios of the neutralino annihilation
 in the $(m_0,m_{1/2})$ plane} \label{bratio} \end{center}
\end{figure}

{\it Region I}: for low $m_{0}$ ($<500-1000$ GeV), the neutralino is almost
purely bino-like. Despite a tiny higgsino fraction, its annihilation via
the $Z$ exchange amplitude, being proportional to the fermion mass, is
surprisingly important in the top channel $\chi\chi\xrightarrow{Z}t\bar{t}
\varpropto (m_tm_{\chi}/m^2_Z)^2N^2_{3(4)}$. Another potentially important
annihilation process is $\chi\chi\xrightarrow{A} b\bar{b}$, whose amplitude
$|A^A_{b\bar{b}}| \varpropto \frac{\tan{\beta}m_{b}}{m_{W}}
(\frac{m_{\chi}}{m_A})^2 N_1N_{3(4)}$ is strongly enhanced for high
$\tan{\beta}$, both from the explicit dependence, and because $m_A$
decreases when $\tan{\beta}$ increases. In the region of interest, their
ratio $|A^Z_{t\bar{t}}/A^A_{b\bar{b}}|$ accordingly drops from $\sim 20-50$
for $\tan\beta=10$ down to $\sim 2-5$ for $\tan\beta=45$. Even taking the
smaller $t\bar{t}$ phase space (a factor of $0.7$) and the other
subleading $b\bar{b}$ processes into account, the $b\bar{b}$ domination
thus calls for an explanation on figure \ref{bratio} and even more on figure
\ref{bratio10}. The only way to account for this domination is to recall
the opposite sign of the $\chi\chi\xrightarrow{Z} t\bar{t}$ and
$\chi\chi\xrightarrow{\tilde{t}} t\bar{t}$ amplitudes, thus allowing for a
cancellation provided their order of magnitude match, which we checked
analytically. The opposite happens for the annihilation into $b\bar{b}$:
the $Z$, $A$ and sbottom exchange all add up, and from the behaviour of the
non-negligible $A^A_{b\bar{b}}$ amplitude, we expect in the region I an
important total annihilation cross section $\sigma^A_{\chi-\chi}$ for high
values of $\tan{\beta}$.

{\it Region II}: when $m_0$ increases ($<1500-2000$ GeV), $m_A$ increases
and, for intermediate $\tan{\beta}$ values, the
$\chi\chi\xrightarrow{A}b\bar{b}$ amplitude decreases. Since sfermion
masses also increase, the $\chi\chi\xrightarrow{\tilde{t}} t\bar{t}$
channel no longer cancels $\chi\chi\xrightarrow{Z} t\bar{t}$, which becomes
the dominant process. However, for high $\tan{\beta}$,
$\chi\chi\xrightarrow{A} b\bar{b}$ is enhanced and stays
dominant. Globally, since all scalar masses have been increased by $m_0$,
and the higgsino fraction is still small, the neutralino annihilation cross
section $\sigma^A_{\chi-\chi}$ is smaller than in region I.

{\it Region III}: increasing further $m_{0}$ ($>2000-2500$ GeV), for any
value of $\tan{\beta}$, one finally approaches the mixed higgsino-bino
region. The $\chi\chi\xrightarrow{Z} t\bar{t}$ channel amplitude
$A^Z_{t\bar{t}}\varpropto(\frac{m_tm_{\chi}}{m^2_Z})N^2_{3(4)}$ then
dominates all other processes, which are suppressed by the increase of
scalar masses. So we are left with a $t\bar{t}$ region parallel to the
highest higgsino fraction isocurves. The $m_0$ value separating this region
from the previous $\chi\chi\xrightarrow{A} b\bar{b}$ region depends on
$\tan{\beta}$. All in all, the neutralino annihilation cross section
$\sigma^A_{\chi-\chi}$ is strongly enhanced by $Z$ exchange near the
no-EWSB boundary.

{\bf Between top and W/Z thresholds:}

{\it Region IV}: since $m_{\chi}< m_{top}$, the neutralino annihilates
almost only into $b\bar{b}$. Even if some $\chi\chi\xrightarrow{\tilde{b}}
b\bar{b}$ occurs, the $\chi\chi\xrightarrow{A} b\bar{b}$ amplitude is
dominant.

{\it Region V}: increasing $m_0$ (and $m_A$), $\chi\chi\xrightarrow{A}
b\bar{b}$ remains dominant and this process is still quite efficient for
high values of $\tan{\beta}$. For intermediate values of $\tan{\beta}$,
$\chi\chi\xrightarrow{A} b\bar{b}$ and $\chi\chi\xrightarrow{Z} b\bar{b}$
both occur but their amplitudes are small and the total annihilation
$\sigma^A_{\chi-\chi}$ is not efficient.

{\it Region VI}: increasing further $m_0$ disfavours scalar exchange, but
even small, the higgsino fraction allows $\chi\chi\xrightarrow{\chi^+_i}
W^+W^-$ and $\chi\chi\xrightarrow{\chi_i} ZZ$ to dominate and enhance the
total annihilation cross section $\sigma^A_{\chi-\chi}$. Again, the $m_0$
values delimiting the boundary with region V depend on $\tan{\beta}$ (via
$m_A$).

{\bf Below W/Z thresholds:}

{\it Region VII}: the dominant process is
$\chi\chi \rightarrow b\bar{b}$ via $A$ and/or $Z$ exchange, depending on
$m_0$, $\tan{\beta}$ and the higgsino fraction.

This analysis is illustrated on figures \ref{bratio} and \ref{bratio10},
showing the four most important branching ratios for $\tan{\beta}=45$ and
$10$ (the omitted process $\chi\chi\xrightarrow{A}\tau\bar{\tau}$ behaves
as $\chi\chi\xrightarrow{A} b\bar{b}$ but with a smaller amplitude due to
the $m_{\tau}/m_b$ ratio).
\begin{figure}[ht]
 \begin{center} \includegraphics[width=0.5\textwidth]{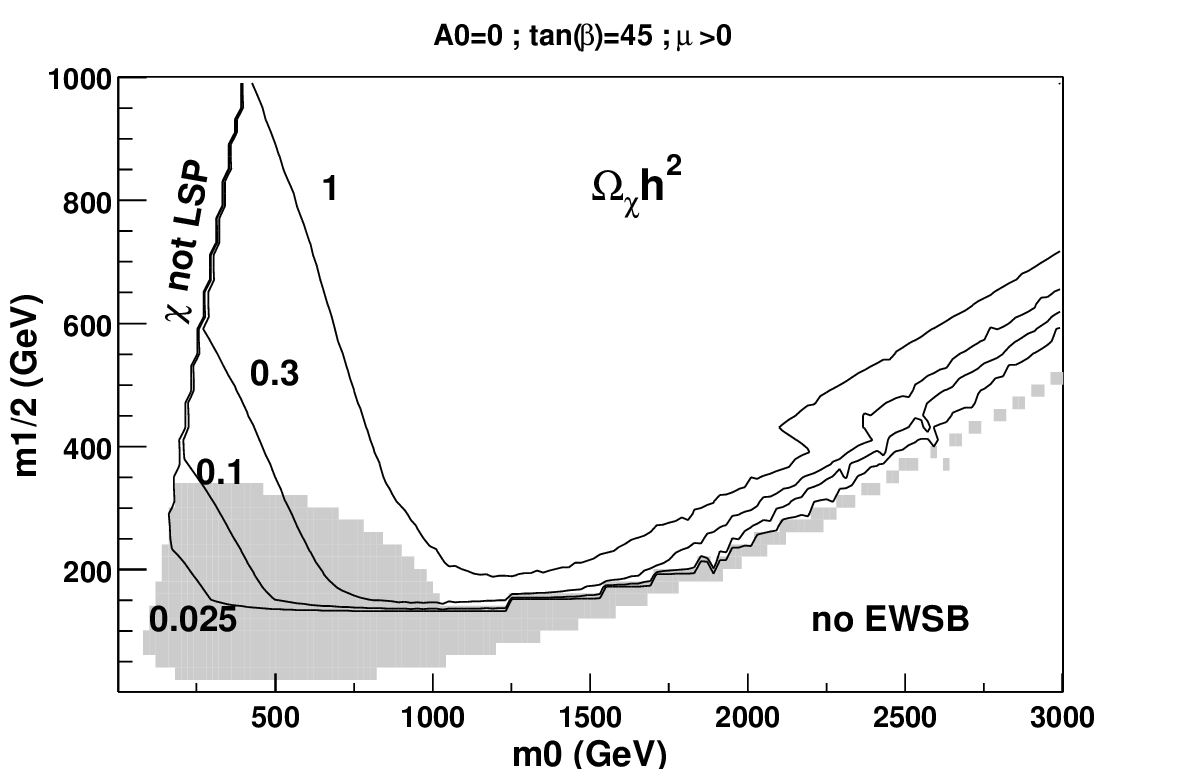}
 \caption{\small Neutralino relic density in the $(m_0,m_{1/2})$
 plane. Grey area indicates the excluded models by current accelerators
 constraints as discussed in the text.}  \label{relic} \end{center}
\end{figure}
It further offers a qualitative understanding of the relic density picture
(figure \ref{relic}). The annihilation is efficient for low values of $m_0$
(depending on $\tan{\beta}$) and for a mixed neutralino along the no-EWSB
boundary. This gives the ``V'' (or ``U'') shape of the relic density
profile for large (or small) $\tan{\beta}$. According to the current
cosmological parameter values, we take the neutralino as an interesting
cold dark matter candidate if $0.025<\Omega h^2<0.3$. Figure \ref{relic}
also shows the region in the $(m_0,m_{1/2})$ plane excluded by the
experimental constraints from the Particle Data Group 2000 \cite{PDG}
implemented in DarkSusy, that we have updated with: \\ $m_{\chi_1^+}>104$
GeV; $m_{\tilde{f}}>100$ GeV for
$\tilde{f}=\tilde{t}_1,\tilde{b}_1,\tilde{l}^{\pm},\tilde{\nu}$,
$m_{\tilde{g}}>300$ GeV; $m_{\tilde{q}_{1,2}}>260$ GeV for
$\tilde{q}=\tilde{u},\tilde{d},\tilde{s},\tilde{c}$ and
$-6\times10^{-10}<a_{\mu}(SUSY)<58\times10^{-10}$ \cite{Knecht:2001qf,Czarnecki:2001pv}.

As noted in previous studies \cite{Feng:2000zu,Djouadi:2001yk}, the muon
anomalous moment $a_{\mu}$and $b\rightarrow s \gamma$ branching ratio
constraints strongly favour $\mu>0$, to which we restrict in what
follows. In figure \ref{relic}, the grey tail at large $m_0$ is excluded by
the chargino bound; it directly bites into the region relevant for neutrino
indirect detection. Less relevant is the exclusion of small $m_0$ and
$m_{1/2}$ values, which comes both from the Higgs mass limit and the
$b\rightarrow s \gamma$ branching ratio, as calculated by default in
DS. The range $BR(b\to s\gamma)=1\to4\times 10^{-4}$ chosen by default in
Darksusy may seem too low in view of latest CLEO and Belle measurements
\cite{PDG}. However the leading order calculation \cite{Bertolini:1991if}
implemented underestimates the SM value to $2.4\times10^{-4}$, while next
to leading order corrections give $3.6\times10^{-4}$ \cite{Ciuchini:1998xe}
so that this range should roughly correspond to $2.2\to5.2\times 10^{-4}$,
excluding a bit more than the range $2\to5\times 10^{-4}$ chosen for
instance in \cite{Djouadi:2001yk}. We have checked that replacing the
implemented Higgs limit \cite{Barate:2000na} by an aggressive version of
the latest limit \cite{Heister:2001kr} ($m_h>114$ GeV for all
$\sin{(\beta-\alpha)}$) only excludes a few more points around ($m_0=1000$,
$m_{1/2}=150$) where neutrino fluxes are beyond reach.

In mSugra, $\chi\chi^+ $ and $\chi\chi_2$ coannihilations (included in DS)
happen only in the mixed neutralino region, decreasing further the relic
density.  $\chi\tilde{\tau} $ coannihilation happens for low values of
$\tan{\beta}$, for which there is no mixed region. $\chi\tilde{t} $
coannihilation \cite{Boehm:1999bj,Djouadi:2001yk,Ellis:2001nx} happens for
high values of $A_0$. In both cases, sfermion coannihilations (missing in
DS) are relevant to lower the relic density in regions of large $m_{\chi}$,
which as we will see, are beyond reach of indirect detection. We therefore do
not expect their proper inclusion to change our conclusions.

\section{Neutralino-proton cross section: capture rate and direct detection 
signals}

{\bf Capture:}\\ If present in the halo, relic neutralinos must accumulate
in astrophysical bodies (of mass $M_b$) like the Sun or the Earth. The
capture rate $C$ depends on the neutralino-quark elastic cross section:
$\sigma_{\chi-q}$. Neutralinos being Majorana particles, their vectorial
interaction vanishes and the allowed interactions are scalar (via $ \chi q
\xrightarrow{H,h} \chi q$ in $t$ channel and $ \chi q
\xrightarrow{\tilde{q}} \chi q$ in $s$ channel) and axial (via $\chi q
\xrightarrow{Z} \chi q$ in $t$ channel and $ \chi q \xrightarrow{\tilde{q}}
\chi q$ in $s$ channel). Depending on the spin content of the nuclei $N$
present in the body, scalar and/or axial interactions are
involved. Roughly, $C\sim \frac{\rho_{\chi}}{v_{\chi}} \sum_NM_bf_N
\frac{\sigma_N}{m_{\chi}m_N} <v^2_{esc}>_N
F(v_{\chi},v_{esc},m_{\chi},m_N)$, where $\rho_{\chi}, v_{\chi}$ are the
local neutralino density and velocity, $f_N$ is the density of nucleus $N$
in the body, $\sigma_N$ the nucleus-neutralino elastic cross section,
$v_{esc}$ the escape velocity and $F$ a suppression factor depending on
masses and velocity mismatching.  The neutralino capture is maximized for
$m_{\chi}\sim m_N$ as $\frac{\sigma_N}{m_{\chi}m_N} \sim
\frac{m^2_{r}}{m_{\chi}m_N} \doteq \frac{m_{\chi}m_N}{(m_{\chi}+m_N)^2}$,
and is much more efficient in the Sun than in the Earth as
$M_{\bigodot}>>M_{\bigoplus}$.

\begin{figure}[!ht]
\begin{center}
 \begin{tabular}{cc}
 \includegraphics[width=0.5\textwidth]{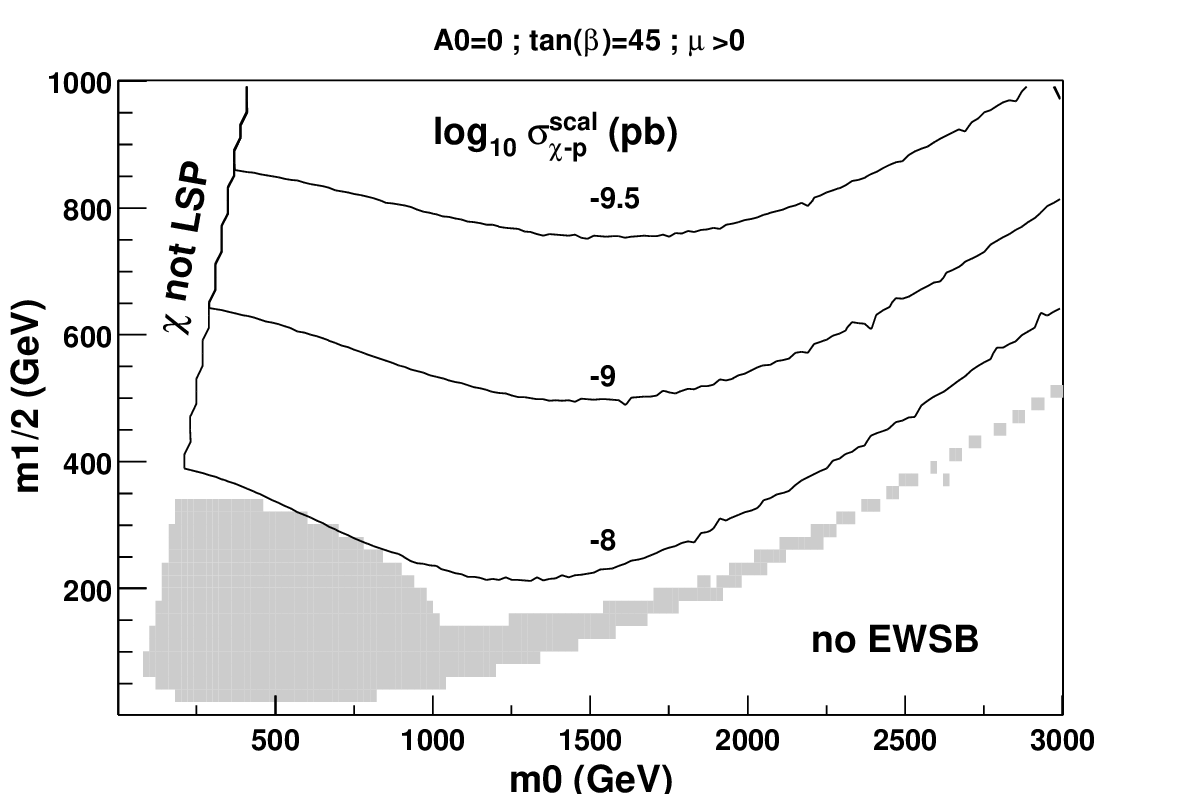} &
\includegraphics[width=0.5\textwidth]{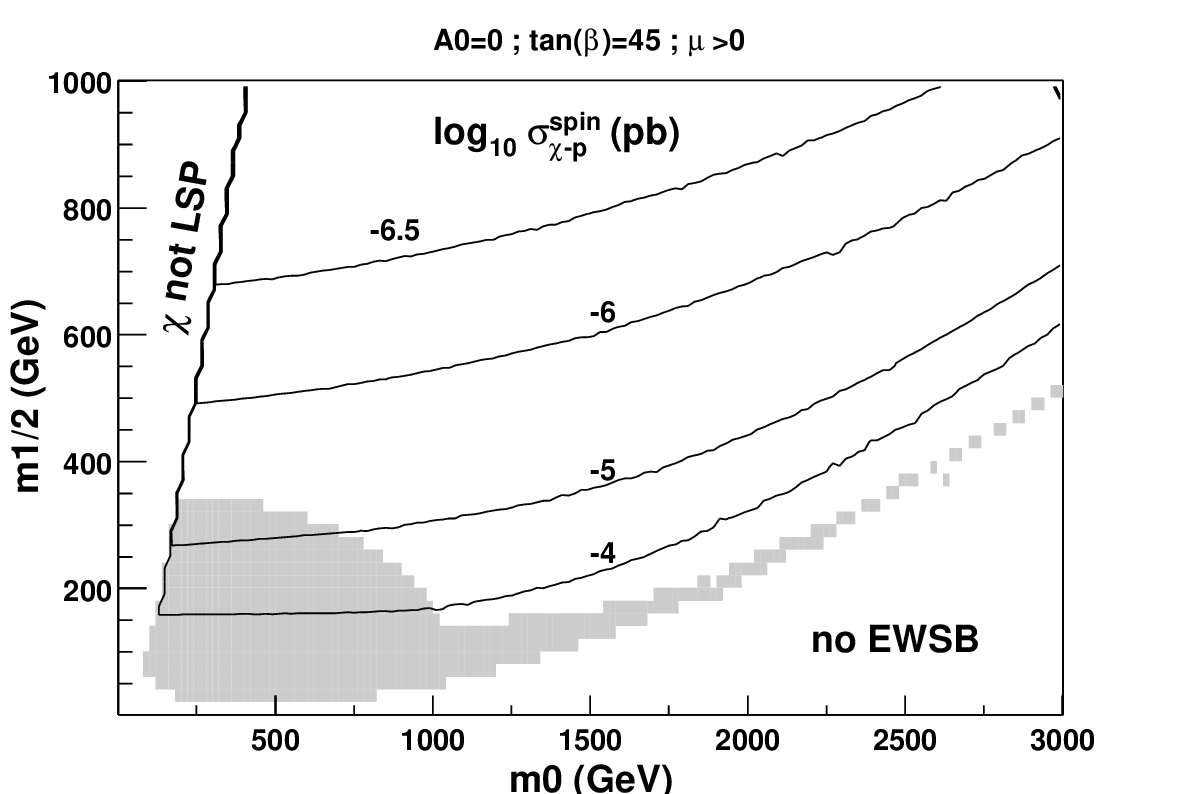}\\
a) & b)
\end{tabular}
 \caption{\small Scalar (a) and  axial (b) cross sections for neutralino
   scattering on proton in pb.}
 \label{sigpsi}
 \end{center}
\end{figure}

{\it For the Earth}, scalar interactions
dominate. By increasing $m_0$ from its low value region, sfermions and $H$
exchanges first decrease, and the cross-section rises again when
approaching the mixed higgsino region (see figure \ref{sigpsi}a). The
capture rate is resonant for $m_{\chi}\sim56 $ GeV around the iron mass.

{\it For the Sun}, the spin of hydrogen allows for axial interaction, which
are stronger due to the $Z$ coupling. The latter depends strongly on the
neutralino higgsino fraction and is independent of $\tan{\beta}$, so the
cross-section follows the higgsino fraction isocurves as can be seen by
comparing figure \ref{sigpsi}b and \ref{isomass}.

To summarize, due to the large solar mass and the spin dependent
neutralino-quark cross section, the storage of neutralinos is much more
efficient in the Sun than in the Earth.

{\bf Direct detection:}\\
The elastic cross section of a neutralino on a nucleus also depends on
$\sigma_{\chi-q}$, the nucleus mass number $A$ and its spin content. Depending
on the chosen nuclei target, current and future direct detection experiments
are sensitive to the scalar coupling (CDMS (Ge) \cite{Abusaidi:2000wg},
Edelweiss (Ge) \cite{Benoit:2002hf,Edelweiss}, Zeplin (Xe) \cite{Zeplin}) or
to the axial coupling (MACHe3 (${\rm^3He}$) \cite{Mayet:2002ke}). Comparison
between experiment sensitivities and a set of mSugra models are shown on
figure \ref{DDandIDandmod}a.

\section{Neutrino indirect detection}

As $\chi\chi\rightarrow \nu\bar{\nu}$ is strongly suppressed by the tiny
neutrino mass, neutrino fluxes come from decays of primary annihilation
products, with a mean energy $E_{\nu}\sim\frac{m_{\chi}}{2}$ to
$\frac{m_{\chi}}{3}$. The most energetic spectra, called ``hard'' come from
neutralino annihilations into $WW$, $ZZ$ and the less energetic, ``soft'',
ones comes from $b\bar{b}$ \cite{Gondolo:2000ee}. Muon neutrinos give rise
to muon fluxes by charged-current interactions in the Earth. As both the
$\nu$ charged-current cross section on nuclei and the produced muon range
are proportional to $E_{\nu}$, high energy neutrinos are easier to
detect. Considering that the population of captured neutralinos has a
velocity below the escape velocity, and therefore neglecting evaporation,
the number $N_{\chi}$ of neutralinos in the centre of a massive
astrophysical object depends on the balance between capture and
annihilation rates: $\dot{N_{\chi}}=C-C_AN_{\chi}^2$, where $C_A$ is the
total annihilation cross section $\sigma^A_{\chi-\chi}$ times relative velocity per volume. The
annihilation rate at a given time $t$ is then:
\begin{equation}
\Gamma_A=\frac{1}{2}C_AN_{\chi}^2=\frac{C}{2}\tanh^2{\sqrt{CC_A}t}
\label{anihirate}
\end{equation}
with $\Gamma_A\approx \frac{C}{2}=cste$ when the neutralino population has
reached equilibrium, and $\Gamma_A\approx \frac{1}{2}C^2C_At^2$ in the
initial collection period (relevant in the Earth). So, when accretion is
efficient, the annihilation rate does not depend on annihilation processes,
but follows the capture rate $C$ and thus the neutralino-quark elastic
cross section.\\
\begin{figure}[!ht]
\begin{center}
\begin{tabular}{c}
 \includegraphics[width=\textwidth]{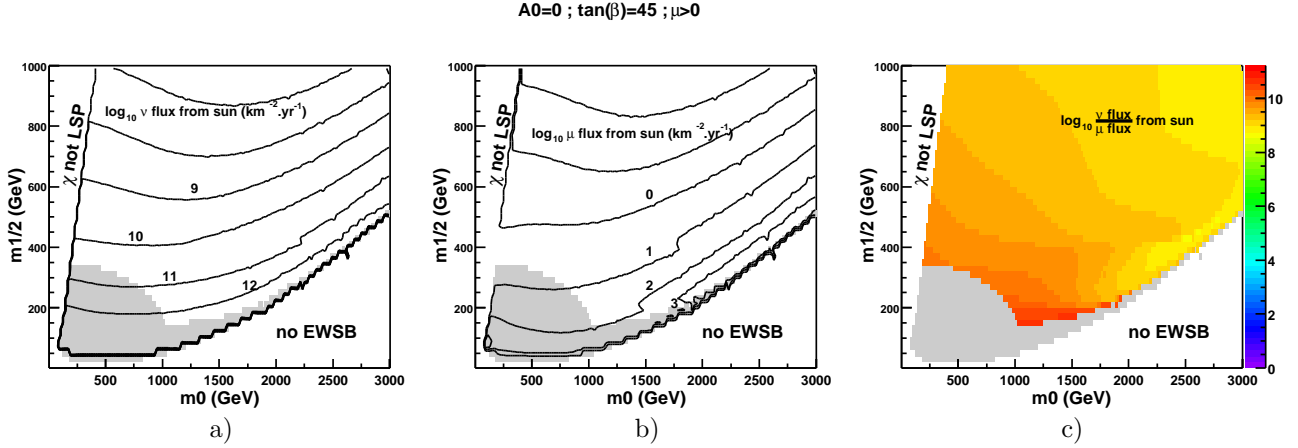}\\
  a) \hspace{0.3\textwidth} b) \hspace{0.3\textwidth} c)
 \end{tabular}
\caption{\small $\nu$ fluxes from the Sun (a), the corresponding $\mu$ fluxes with $E_{\mu}>5$ GeV theshold (b) and their ratio (c).}
 \label{sunflux}
 \end{center}
\end{figure}

{\it For the Sun}, neutralinos do not reach complete equilibrium in the whole
$(m_0,m_{1/2})$ plane studied, which contrasts with \cite{Feng:2000zu} based
on Neutdriver \cite{Jungman:1996df} and different RGE, probably yielding
lighter (pseudo-)scalars and lower $\mu$. We find that the annihilation is on
average less efficient (as attested by our smaller region with an acceptable
relic density), which results in a filling fraction $\sqrt{CC_A}t_{\bigodot}$
smaller by a factor of 10 to 100. This has little effect on $\nu(\mu)$ fluxes
for high $m_0$ or small $m_{1/2}<600$~GeV values, as equilibrium is
nevertheless reached and fluxes follow essentially the higgsino fraction and
the spin dependent $\sigma^{spin}_{\chi-p}$ isocurves of figures
\ref{isomass}a and \ref{sigpsi}. For low $m_0$ however, the equilibrium fluxes
would drop with $m_0$ and $C$, but the smaller capture rate hinders
equilibrium ({\it e.g. }$\sqrt{C C_A}t_{\bigodot}\sim 1$ for $m_0=500$,
$m_{1/2}=800$) and $\Gamma_A$ feels the increasing annihilation cross-section.
This effect is stronger for high $m_{1/2}$ values where the neutralinos are
heavier and more bino-like, and where capture is smaller. The effect is also
larger when $\tan{\beta}$ (and thus $\sigma^A_{\chi-\chi}$) is low, as
incomplete equilibrium makes $\nu(\mu)$ fluxes sensitive to
$\sigma^A_{\chi-\chi}$, making the $W$ and top thresholds more conspicuous.

\begin{figure}[!ht]
\begin{center}
\begin{tabular}{c}
 \includegraphics[width=\textwidth]{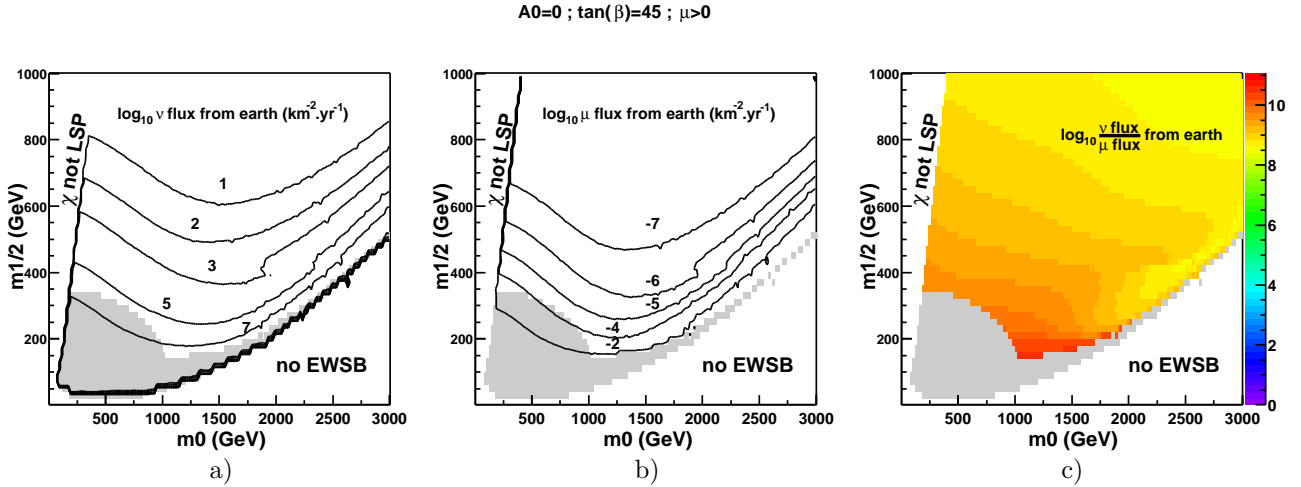}\\
  a) \hspace{0.3\textwidth} b) \hspace{0.3\textwidth} c)
 \end{tabular}
\caption{\small $\nu$ fluxes from the Earth (a), the corresponding $\mu$
fluxes $E_{\mu}>5$ GeV theshold (b) and their ratio (c).}
\label{earthflux} \end{center}
\end{figure}

{\it For the Earth}, neutralinos are not in equilibrium. Neutrino fluxes
depend both on $C^2$ and annihilation, giving an enhancement in the low and
high $m_0$ regions where fluxes are boosted by annihilation (see figure
\ref{sigpsi} and \ref{earthflux}). Since $M_{\bigoplus}<M_{\bigodot}$ and
$\sigma^{scal}_{\chi-p}<\sigma^{spin}_{\chi-p}$, the capture rate and $\nu$
fluxes from the Earth are much smaller than from the Sun.\\

{\it Comparing $\nu$ fluxes and $\mu$ fluxes}, the $\nu\to\mu$ conversion
factor increases with $m_{1/2}$ due to the increase of $m_{\chi}$ leading
to more energetic neutrinos. This ratio also follows the annihilation final
state regions described in section 3 above. Indeed, spectra from $WW$, $ZZ$
and to a lesser extent $t\bar{t}$ are more energetic than $b\bar{b}$
spectra, so neutrino conversion into muon is more efficient in the mixed
bino-higgsino region above $W$ and top thresholds (see figure \ref{sunflux}
and \ref{earthflux}).\\

\begin{figure}[!ht]
\begin{center}
\begin{tabular}{c}
\includegraphics[width=\textwidth]{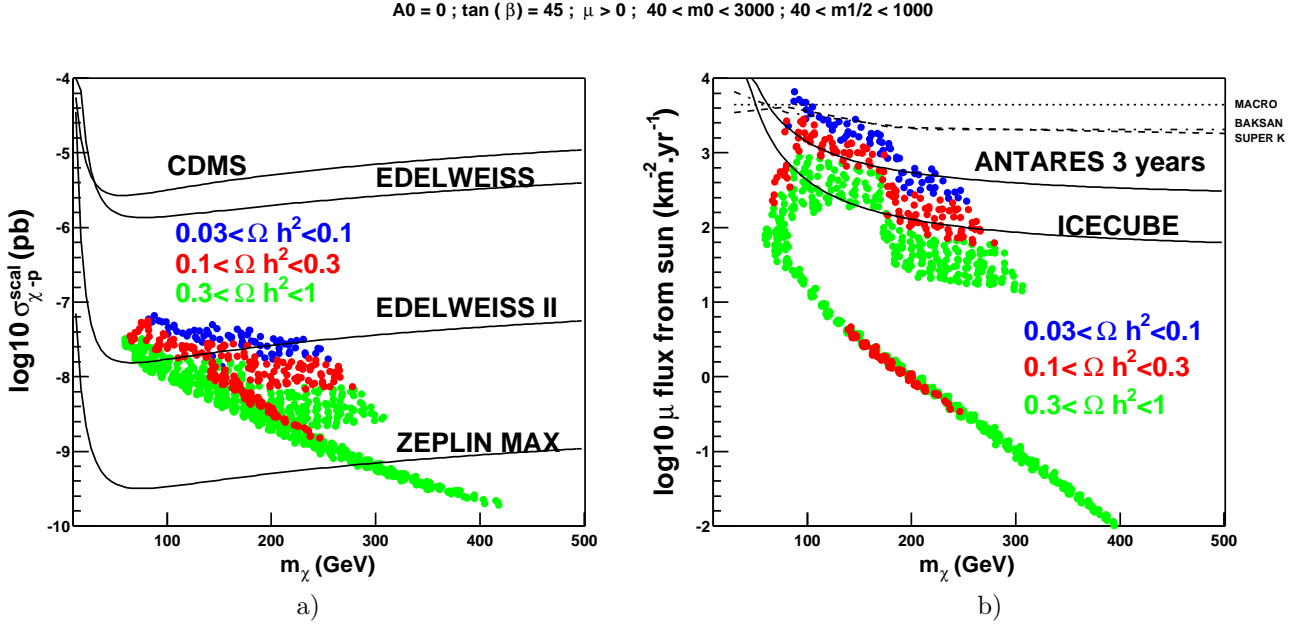}\\
  a) \hspace{0.5\textwidth} b)
 \end{tabular} 
\caption{\small a) Direct detection experiment sensitivities and a set of
models in the $(\sigma^{scal}_{\chi-p},m_{\chi})$ plane. The models
excluded by current accelerators constraints are not displayed. b) Indirect
detection experiment sensitivities in the $(\mu \
flux_{\bigodot},m_{\chi})$ plane with $E_{\mu}=5$ GeV threshold of the same
sample as a). Dotted, dash-dotted and dashed curves are respectively the
Macro, Baksan and Super-Kamiokande upper limits. Solid lines indicate the
future Antares and Icecube sensitivities.}  \label{DDandIDandmod}
\end{center}
\end{figure}
   {\it Comparison with indirect detection experiment sensitivities:} Muon
   fluxes coming from neutralino annihilation in the Sun are shown in figure
   \ref{DDandIDandmod}b. MSugra models predicting a good relic density of
   neutralinos give neutrino/muon fluxes which can be as high as the current
   experimental limits (Baksan \cite{Suvorova:1999my}, Macro \cite{Macro},
   Super Kamiokande \cite{SuperK}). The $0.1 \ {\rm km^2}$ Antares detector
   will explore further the interesting parameter space \cite{AntarLee}. Next
   generation neutrinos telescopes (Icecube, Antares ${\rm km^3}$) will be
   much more efficient to test such models, eg Icecube expected sensitivity
   $\sim10^2\ \mu\ {\rm km^2 \ yr^{-1}}$ from the Sun \cite{Ice3Edsjo}.\\

\begin{figure}[!ht]
\begin{center}
\begin{tabular}{c}
 \includegraphics[width=\textwidth]{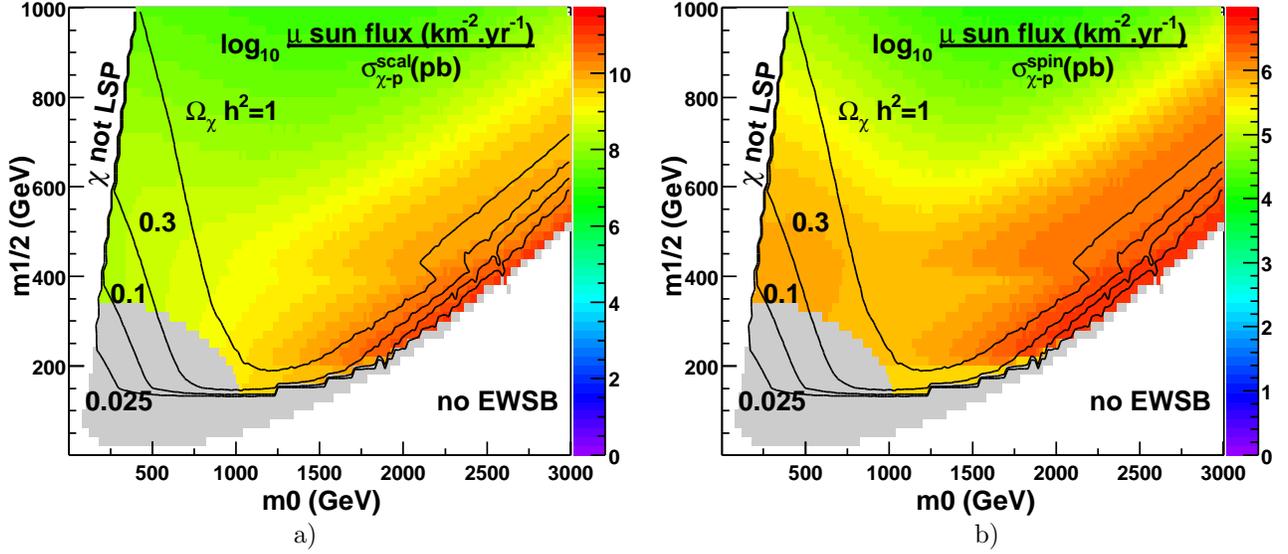}\\ a)
  \hspace{0.5\textwidth} b) \end{tabular} \caption{\small Comparison
  between $\mu$ fluxes from the Sun and neutralino-proton cross sections:
  a) scalar, b) spin dependent} \label{indirect-on-direct} \end{center}
\end{figure}

{\it Indirect vs Direct detection:} a high neutralino-proton cross section
is efficient in both direct and indirect detection (via capture). The large
$m_0$ mixed higgsino-bino region pointed out in
\cite{Feng:2000gh,Feng:2000zu} favours both direct and indirect detection
due to the enhancement of $\sigma_{\chi-p}$ and
$\sigma^A_{\chi-\chi}$. Both enter in the indirect detection which is
moreover favoured by the production of more energetic neutrinos in $WW,\
ZZ$ and $t\bar{t}$ decays, leading to better conversion into muons (figure
\ref{indirect-on-direct}). This mixed region, which has a good relic
density, is very attractive for neutrino indirect detection signal.

\begin{figure}[htb]
\begin{center}
\begin{tabular}{c}
 \includegraphics[width=\textwidth]{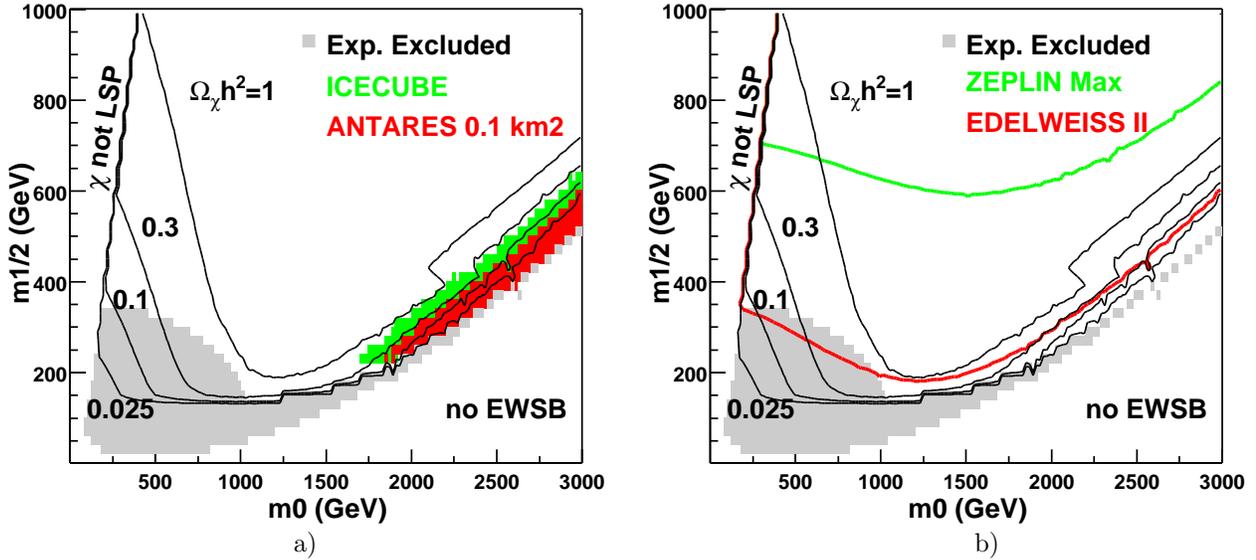}\\ a)
  \hspace{0.5\textwidth} b) \end{tabular} \caption{\small a) $\nu$
  telescopes sensitivities on $\mu$ fluxes from the Sun and b) direct
  detection experiments sensitivities in the $(m_0,m_{1/2})$ plane.}
  \label{manip} \end{center}
\end{figure}

Hovewer, no current experiment is able to really test such models. Figure
\ref{manip} shows the region of the $(m_0,m_{1/2})$ plane which can be
explored by future direct detection experiments and neutrino telescopes.\\

\section{Variations}
In this section, we discuss the robutness of the indirect detection picture
described above, with respect to variations in Suspect 2.002 of 1) mSugra
input parameters like $A_0$ or $\tan{\beta}$, 2) experimental uncertainties
on the top quark mass and 3) theoretical uncertainties on the scale of
electroweak symmetry breaking $Q_{EWSB}$

\begin{figure}[htbp]
\begin{center}
\begin{tabular}{c}
 \includegraphics[width=\textwidth]{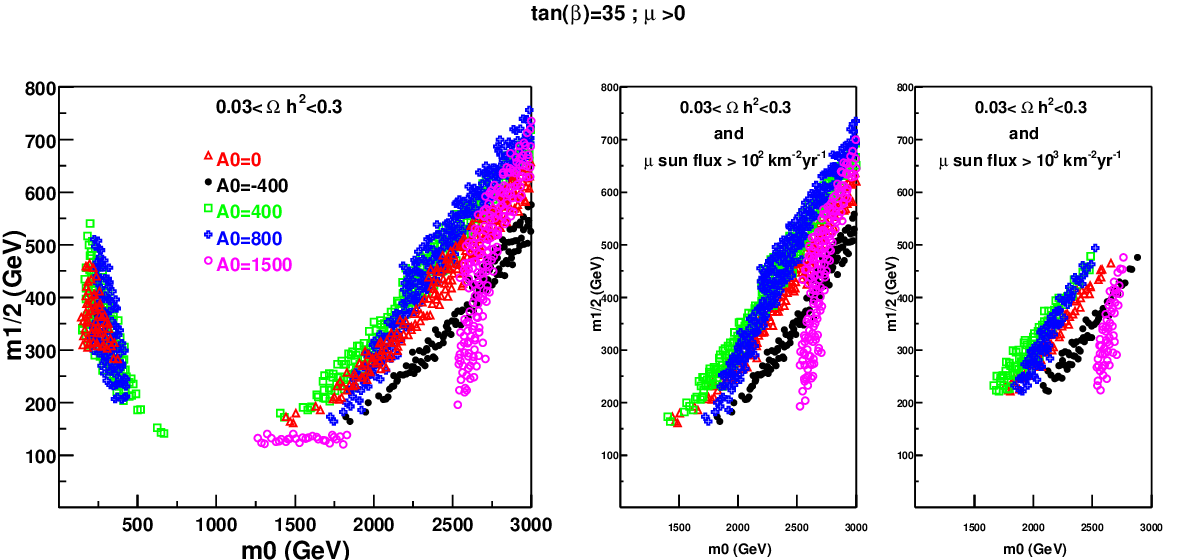}\\
a)\hspace{0.45\textwidth} b)\hspace{0.25\textwidth} c) \end{tabular}
\caption{\small a) Cosmologically favoured models for differents values of
$A_0$ in the $(m_0,m_{1/2})$ plane; subsetof these models generating a
$\mu$ flux from the Sun larger than b) $10^2 \ {\rm km^{-2} \ yr^{-1}}$; c)
$10^3 \ {\rm km^{-2} \ yr^{-1}}$.}  \label{A0} \end{center}
\end{figure}

{\bf ${\bf A_0}$:} Varying $A_0$ away from $0$ does not change too much the
analysis above. Annihilation branching ratio regions, as well as the
no-EWSB boundary and slope can be slightly displaced. For moderate
$\tan{\beta}\sim20$, quite large and negative values of $A_0$ (\emph{eg}
$-800$ GeV) can enhance $\tau^+\tau^-$ annihilation channel along the
$\tilde{\tau}$ LSP region (\emph{i.e.} low $m_0$) due to splitting in the
$\tilde{\tau}$ mass matrix. For such values, the $\tilde{\tau}\chi$
coannihilation \cite{Ellis:1999mm} could also affect the relic density, but
in this region, neutrino fluxes are too small to be detected anyway (see
figure \ref{A0}). The same happens with the $t\bar{t}$ channel for
$400<m_{1/2}<600$ and $m_0<700 \ (\sim|A_0|)$. The $\tilde{t}\chi$
\cite{Boehm:1999bj,Djouadi:2001yk,Ellis:2001nx} coannihilation can also
happen for $A_0\sim-2000$ GeV, but the usual cosmologically interesting
region is reduced (due to the negative squared mass arising in sfermion
matrices at low $m_0$ and because the mixed region is pushed to higher
values of $m_0$). In addition, this coannihilation region produces neutrino
fluxes from the Sun around $10^{-1} \mu \ {\rm km^{-2} \ yr^{-1}}$, too low
to be detected. In view of the latest results on cosmological parameters,
figure \ref{A0}a shows models with $0.03<\Omega h^2<0.3$ in the
$(m_0,m_{1/2})$ plane for different values of $A_0$, as well as models
giving $\mu$ fluxes from the Sun larger than $10^2$ (b) and $10^3 \ \mu \
{\rm km^{-2} \ yr^{-1}}$ (c). Latter models are confined in the mixed
higgsino-bino region. In addition, only those models with
$m_W<m_{\chi}<m_t$ giving a hard spectrum of neutrinos via
$\chi\chi\xrightarrow{\chi^+_i} W^+W^-$ and
$\chi\chi\xrightarrow{\chi_i}ZZ$ can simultaneously satisfy $0.1<\Omega
h^2<0.3$, while yielding a high muon flux.\\

\begin{figure}[htbp]
\begin{center}
 \includegraphics[width=\textwidth]{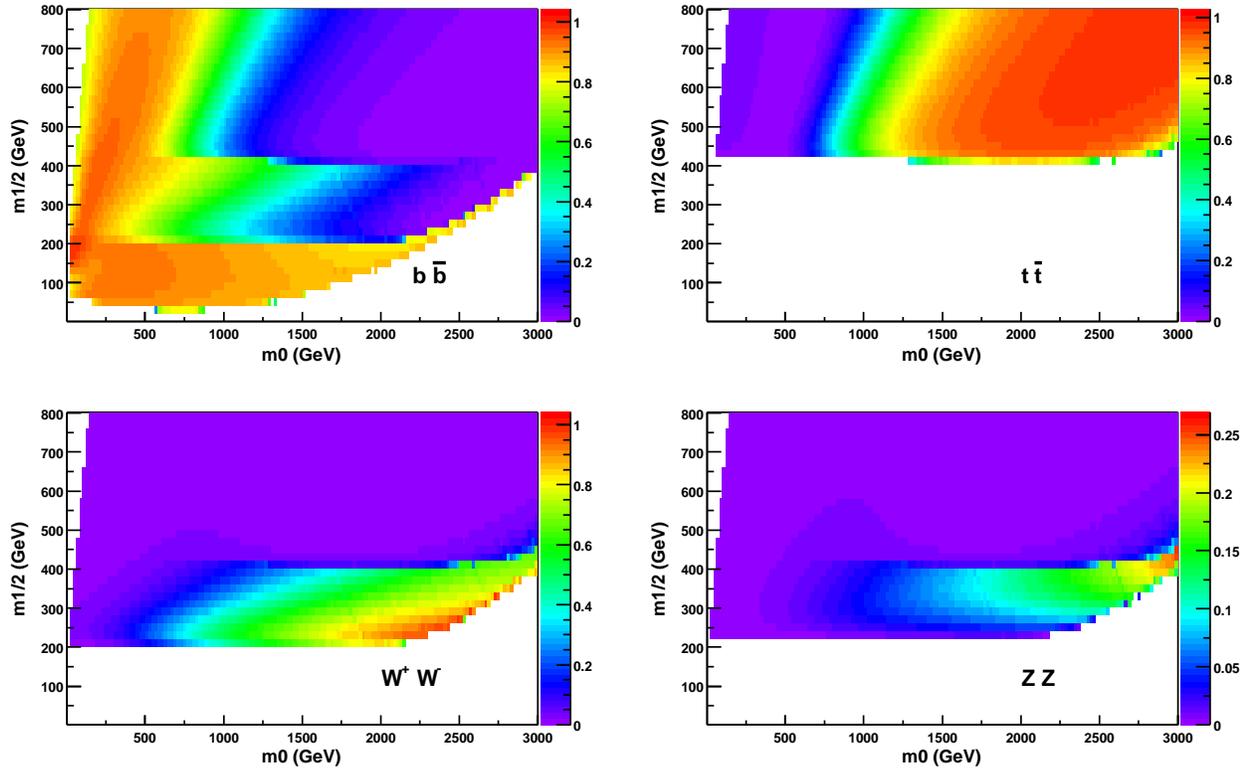} 
 \caption{\small
 Branching ratio of the neutralino annihilation in the $(m_0,m_{1/2})$
 plane (same as figure \ref{bratio} for $\tan{\beta}=10$).}
 \label{bratio10} \end{center}
\end{figure}

\begin{figure}[htbp]
\begin{center}
\begin{tabular}{c}
 \includegraphics[width=\textwidth]{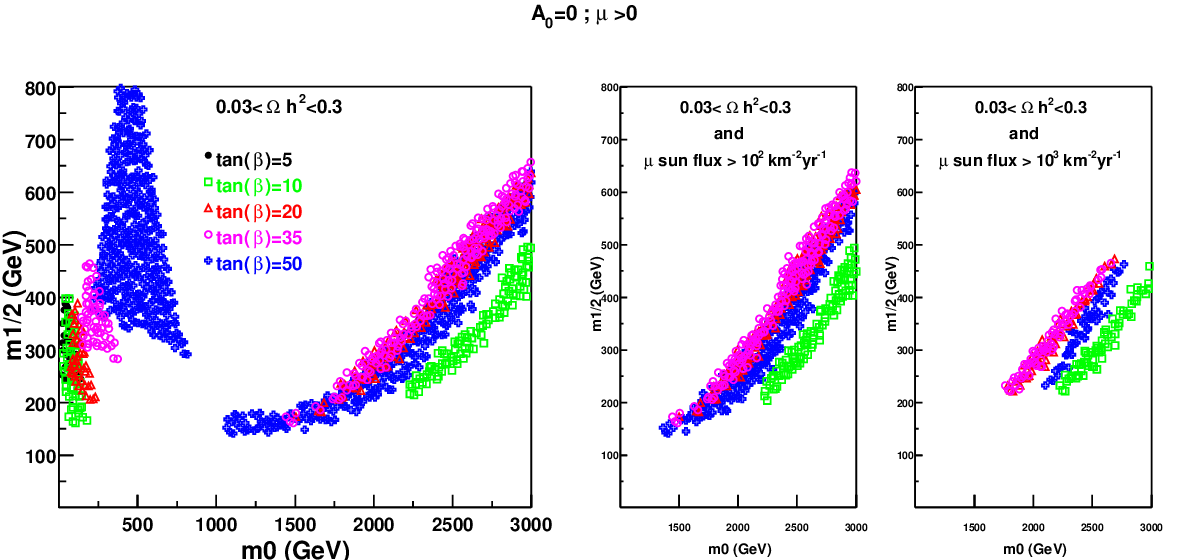}\\
a)\hspace{0.45\textwidth} b)\hspace{0.25\textwidth} c) \end{tabular}
\caption{\small a) Cosmologically favoured models for differents values of
$\tan{\beta}$ in the $(m_0,m_{1/2})$ plane; subsetof these models
generating a $\mu$ flux from the Sun larger than b) $10^2 \ {\rm km^{-2} \
yr^{-1}}$; c) $10^3 \ {\rm km^{-2} \ yr^{-1}}$.}  \label{tanbeta}
\end{center}
\end{figure}

{\bf tan $\beta$:} the branching ratio picture does not qualitatively
change for different values of $\tan{\beta}$. Low values of $\tan{\beta}$
can add a $\chi\chi\rightarrow t\bar{t}$ region for low $m_0$ and open a
$\tilde{\tau}\chi$ coannihilation \cite{Ellis:1999mm} region along the
$\tilde{\tau}$ LSP region, but $\mu$ fluxes are small and such values of
$\tan{\beta}$ reduce the mixed neutralino region which is the most
interesting for indirect detection. As discussed previously, the boundaries
of branching ratio regions move with $\tan{\beta}$ (compare figure
\ref{bratio10} and \ref{bratio}). When $\tan{\beta}$ grows, the
cosmologically motivated region is wider due to the annihilation
enhancement via $A$ exchange (in regions I/II and IV/V). Applying the
conservative cut $0.03<\Omega h^2<0.3$, all values of $\tan{\beta}$ provide
interesting models for neutrino indirect detection (see figure
\ref{tanbeta}). For too large values of $\tan{\beta}$, the $b\bar{b}$
region is extended and leads to soft neutrino spectra. In this case it
becomes more difficult to conciliate a tighter limit on the relic density
with current experimental sensitivities. Values of $\tan{\beta}\sim 10 \to
35$ give more models satisfying both relic density $0.1<\Omega h^2<0.3$ and
high $\nu/\mu$fluxes.\\

\begin{figure}[!ht]
\begin{center}
 \begin{tabular}{c}
 \includegraphics[width=\textwidth]{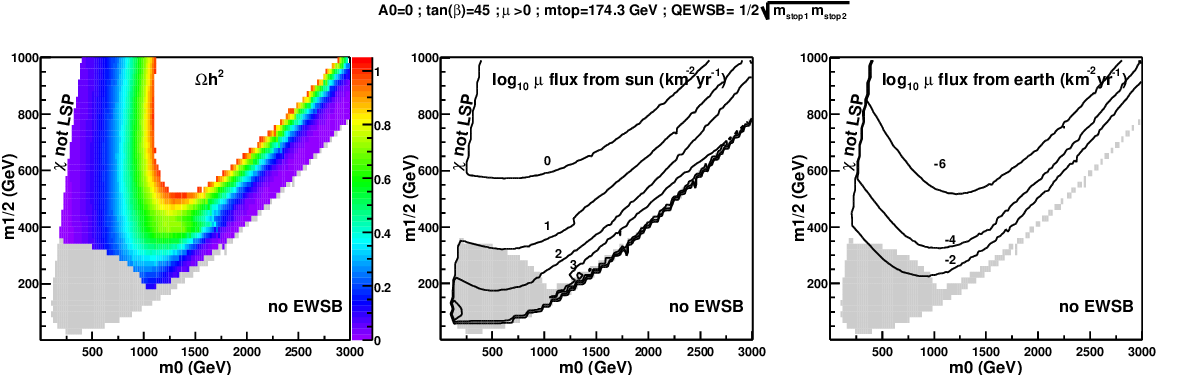}\\ a)
 \hspace{0.3\textwidth} b) \hspace{0.3\textwidth} c) \end{tabular}
 \caption{\small a) Relic density, $\mu$ fluxes from the Sun (b) and the
 Earth (c) in the $(m_0,m_{1/2})$ plane for
 $Q_{EWSB}=\frac{1}{2}\sqrt{m_{\tilde{t}_1}m_{\tilde{t}_2}}$ and
 $m_t=174.3$ GeV.}  \label{EWSB} \end{center}
\end{figure}

\begin{figure}[!ht]
\begin{center}
 \begin{tabular}{c}
	\includegraphics[width=\textwidth]{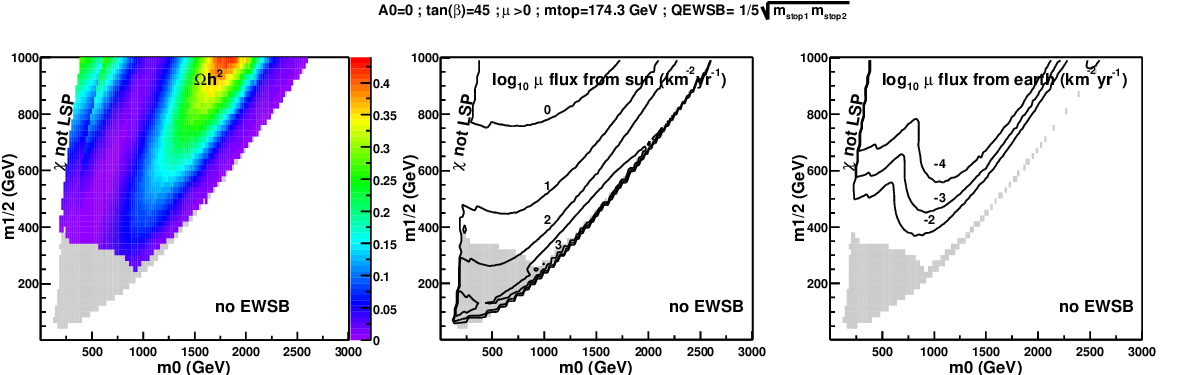}\\ 
	a) \hspace{0.3\textwidth} b) \hspace{0.3\textwidth} c) 
 \end{tabular}
 \caption{\small a) Relic density, $\mu$ fluxes from the Sun (b) and the
 Earth (c) in the $(m_0,m_{1/2})$ plane for $Q_{EWSB} = \frac{1}{5}
 \sqrt{m_{\tilde{t}_1}m_{\tilde{t}_2}}$ and $m_t=174.3$ GeV.}
 \label{1on5EWSB} \end{center}
\end{figure}

\begin{figure}[!ht]
\begin{center}
 \begin{tabular}{c} \includegraphics[width=\textwidth]{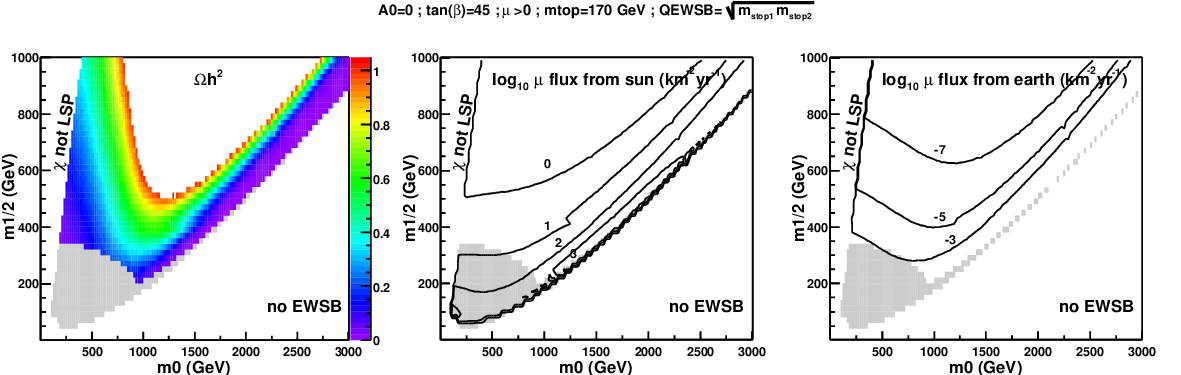}\\
	 a) \hspace{0.3\textwidth} b) \hspace{0.3\textwidth} c)
	\end{tabular}
 \caption{\small a) Relic density, $\mu$ fluxes from the Sun (b) and the
 Earth (c) in the $(m_0,m_{1/2})$ plane for $Q_{EWSB} =
 \sqrt{m_{\tilde{t}_1} m_{\tilde{t}_2}}$ and $m_t=170$ GeV.}
 \label{EWSB170} \end{center}
\end{figure}

\begin{figure}[!ht]
\begin{center}
 \begin{tabular}{c}
 \includegraphics[width=\textwidth]{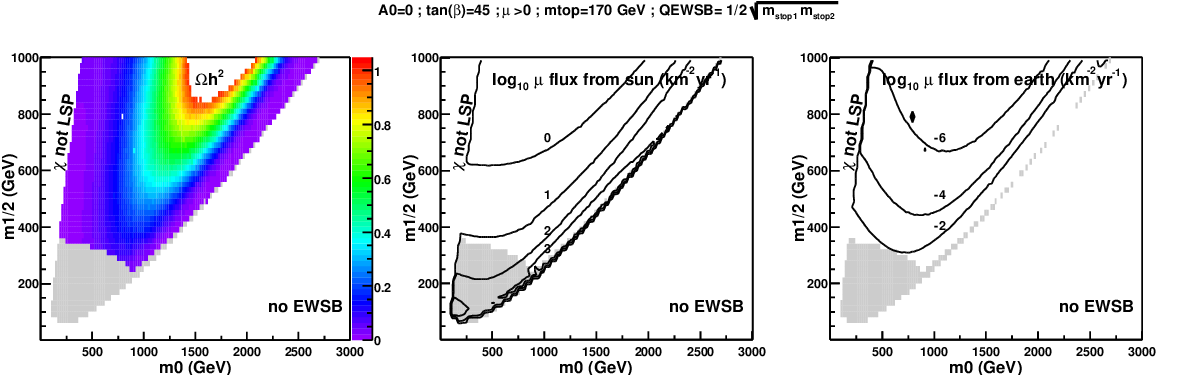}\\
 a) \hspace{0.3\textwidth} b) \hspace{0.3\textwidth} c)
 \end{tabular}
 \caption{\small a) Relic density, $\mu$ fluxes from the Sun (b) and the Earth
 (c) in the $(m_0,m_{1/2})$ plane for $Q_{EWSB}=\frac{1}{2} \sqrt{m_{\tilde{t}_1}m_{\tilde{t}_2}}$ for $m_t=170$ GeV.}
 \label{halfEWSB170}
 \end{center}
\end{figure}

\begin{figure}[htbp]
\begin{center}
\begin{tabular}{c}
 \includegraphics[width=\textwidth]{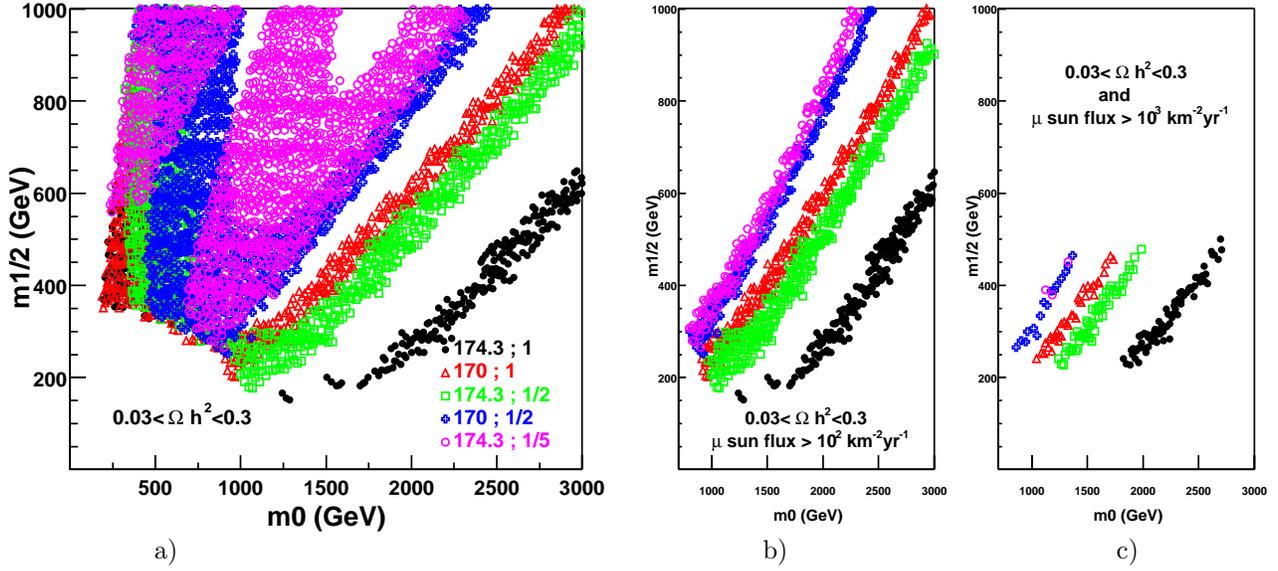}\\
	a)\hspace{0.45\textwidth} b)\hspace{0.25\textwidth} c)
	\end{tabular} \caption{\small a) Dark matter favoured models for
	differents values of $Q_{EWSB}$ and top mass in the $(m_0,m_{1/2})$
	plane (a). Colors and symbols correspond to different values of
	$m_t$ and $Q_{EWSB}$: \emph{e.g.}  (170 ; 1/2) means $m_t=170$ GeV
	and $Q_{EWSB} = \frac{1}{2} \sqrt{m_{\tilde{t}_1}
	m_{\tilde{t}_2}}$. The models that moreover yield a muon flux from
	the Sun larger than $>10^2 \ {\rm km^{-2} \ yr^{-1}}$ and $>10^3 \
	{\rm km^{-2} \ yr^{-1}}$ are shown in b) and c).}
\label{cumul-EWSBmtop}
\end{center}
\end{figure}

{\bf EWSB scale:} $Q_{EWSB}$ is the scale at which the running of the soft
SUSY breaking terms is frozen and the minimization of the scalar potential
is achieved. When the full potential is minimized within a multi-scale RG
treatment, $Q_{EWSB}$ should not affect physics. Hovewer, the one-loop
single scale effective potential does depend on $Q_{EWSB}$, and the
tree-level relations (equation \ref{potminimi}) only hold in terms of
running parameters at $Q_{EWSB}$; if $Q_{EWSB}$ is chosen
$\sim\sqrt{m_{\tilde{t}_1}m_{\tilde{t}_2}}$ \cite{Gamberini:1990jw}, the
potential can be minimized pertubatively. Varying $Q_{EWSB}$ away from this
value is therefore not physical, but gives a hint on the theoretical
uncertainties associated with a single scale RG and potential minimization
approach. When $Q_{EWSB}$ is lowered, the RG running drives $\mu$ to
smaller values. This leads to a wider region of the $(m_0,m_{1/2})$ plane
where EWSB cannot be achieved. In the remaining allowed region, the lower
values of $\mu$ slightly increases the higgsino fraction of the neutralino
and the annihilation cross-section, giving more cosmologically acceptable
models. Figure \ref{EWSB} and \ref{1on5EWSB} show the relic density and
$\mu$ fluxes from the Sun and the Earth in the $(m_0,m_{1/2})$ plane for
$Q_{EWSB}=\frac{1}{2}\sqrt{m_{\tilde{t}_1}m_{\tilde{t}_1}}$ and
$\frac{1}{5}\sqrt{m_{\tilde{t}_1}m_{\tilde{t}_1}}$. For
$Q_{EWSB}=\frac{1}{5}\sqrt{m_{\tilde{t}_1}m_{\tilde{t}_2}}$, one sees
around $m_0=m_{1/2}$ the $\chi\chi\xrightarrow{A} b\bar{b}$ $s-$channel
annihilation pole which would otherwise show up for higher values of
$\tan{\beta}$ $\gtrsim55$ \cite{Djouadi:2001yk}. It affects significantly the fluxes from the
Earth, where capture and annihilation are not in equilibrium figure
\ref{1on5EWSB}c. The higher higgsino fraction also enhances the elastic
cross section $\sigma_{\chi-q}$, but the stronger annihilation reduces the
number of models which are both interesting from the cosmological point of
view and in terms of Sun muon fluxes (figure \ref{cumul-EWSBmtop}c). As
before, muon fluxes from the Earth are much too weak to be detected.
Increasing $Q_{EWSB}$ leads to the opposite effect: higher values of
$|\mu|$ and reduced cosmological favoured region.

{\bf Top mass:} the current experimental value is $m_t=174.3 \pm 5.1$ GeV
\cite{PDG}. Our default choice up to now has been to adopt the central
value. Lowering $m_t$ implies that the top Yukawa coupling $Y_t$ is smaller
so that the RG decreasing of $m^2_{H_u}$ is less efficient and the region
where EWSB does not occur is wider (see eq. \ref{potminimi}, figure
\ref{EWSB170} and \ref{halfEWSB170}). Nevertheless the mixed higgsino
region is larger favouring annihilation and capture, but there are no
additional models combining a large $\nu/\mu$ flux with an acceptable relic
density (figure \ref{EWSB170}, \ref{halfEWSB170} and
\ref{cumul-EWSBmtop}). Increasing the top mass leads to the opposite
effect, which is less interesting for our study. It should also be noticed
that varying $Q_{EWSB}$ and/or $m_t$ allows for heavier neutralino within
the fixed $\Omega h^2$ interval.

\section{Conclusion}
In this paper, we have analysed, within the framework of the constrained
Minimal Supersymmetric Standard Model (\emph{a.k.a.} mSugra), the possible
indirect detection of neutralino dark matter through the neutrinos produced
by its annihilation in the Earth and the Sun. In particular, we have
presented an original study of the relative weight of various annihilation
channels crucial to indirect detection. Fixing the halo dark matter density
to $\rho_{\chi}=0.3 \ {\rm GeV}/{\rm cm}^3$ and allowing for the widest
range of cosmological relic density \(0.03<\Omega h^2<0.3\), we find that a
neutrino signal from the centre of the Earth is beyond reach of the present
and even future neutrino telescopes.

For the larger Sun, we find that a neutrino signal may be found, but only
in the large $m_0$ "focus point" region \cite{Feng:2000gh,Feng:2000zu}
where the neutralino has a larger higgsino component which increases both
neutralino-nucleon elastic scattering and neutralino annihilation
cross-sections. Since in this region, both are related by $s$- and
$t$-channel crossing (exactly above $t\bar t$ threshold and approximately
below), any parameter change reducing the relic density inevitably
increases the neutrino flux. When conversion into muons is taken into
account with a 5 GeV threshold on the muon energy, the only models
providing both a relevant (but rather low) relic density, and fluxes around
the {\it current} indirect detection sensitivity ($10^3 \mu\ {\rm km}^{-2}
\ {\rm yr}^{-1}$) are those with $m_\chi<m_t$ and $\chi\chi \xrightarrow
{\chi^+_i} W^+W^-$ or $\chi\chi\xrightarrow{\chi_i} ZZ$ as dominant
annihilation processes. Relaxing the tight mSugra constraints in specific
directions could favorise such channels \cite{mynext}. The next generation
of neutrino telescopes (with a ${\rm km}^3$ size) will be much more
efficient to probe mSugra models, especially for $m_\chi>m_t$. If a
neutrino signal is found in this framework, then neutralino dark matter
should also be accessible to future direct detection experiments, and a
chargino with $m_{\chi^+}<350 \ {\rm GeV}$ should be found in accelerators.

To conclude, a comparison with previous related work is in order. Earlier
studies \cite{Jungman:1996df,Bergstrom:1998xh} of neutralino indirect
detection were performed in the unconstrained MSSM, where $\mu$ is a free
parameter, and mixed neutralinos occur rather easily. Taking an mSugra
slice in this huge parameter space obviously introduces many correlations,
for instance between the neutrino flux and the relic density or
$\sigma_{\chi-p}^{spin}$ (and even $\sigma_{\chi-p}^{scal}$). The effect of
this slicing in the $(m_\chi,\sigma_{\chi-p}^{scal})$ plane was shown in
\cite{Ellis:2000ds} for small $\tan\beta$ and in \cite{Ellis:2001qm} for
large $\tan\beta$. Our figure \ref{DDandIDandmod} agrees with this last
result, with the addition of the focus point region forming the upper-right
cloud. We should note however that a comparison in the $(m_0,m_{1/2})$
plane is more difficult, specially at large $\tan\beta$, because of the
theoretical uncertainties in the RGE used to translate these parameters
into physical quantities. As an example, for a given value of $\tan\beta$,
the slope of the EWSB boundary decreases between \cite{Feng:2000zu}, this
work and \cite{Ellis:2001qm}. As another example, the strong annihilation
via $s$-channel $A,H_0$ scalars appears above $\tan\beta>50$ in
\cite{Barger:2001ur} and \cite{Roszkowski:2001sb}, $\tan\beta>35$ in
\cite{Ellis:2001qm}, and $\tan\beta>60$ (or $\tan\beta=45$,
$Q_{EWSB}=\sqrt{m_{\tilde t_1}m_{\tilde t_2}}/5$, see figure
\ref{1on5EWSB}) in the present work using SUSPECT. Finally, we find a muon
event rate compatible with \cite{Feng:2000zu}, and larger than
\cite{Barger:2001ur}. Part of this difference comes from a higher threshold
(25 GeV instead of our 5 GeV). The rest might be attributed to a high
sensitivity in the renormalization group equations at large $m_0$ and
$\tan\beta$. It seems \cite{Allanach:2001hm} that using SOFTSUSY
\cite{Allanach:2001kg} or SUSPECT \cite{Suspect,Djouadi:2001yk} may be
safer than ISASUGRA\cite{Baer:1999sp} in this region.

{\bf Acknowledgement}
 This work would not have started without the french "GDR supersymetrie". We
 gratefully thank Jean-Loic Kneur, Charling Tao, and the Antares Neutralino WG
 for help and stimulating discussions. We must also underline the acute reading
 and very constructive role of the referee in the final version of this
 article.

\nocite{}
\bibliography{id-msugra}

\begin{thebibliography}{10}

\bibitem{Peacock}
J.~A. Peacock.
\newblock {\em Cosmological Physics}, chapter~3.
\newblock Cambridge University Press, 1999.

\bibitem{Lineweaver:2001}
C.~H. Lineweaver.
\newblock Cosmological parameters.
\newblock 2001.
\newblock talk presented at COSMO-01, Rovaniemi, Finland August 29 - September
  4.

\bibitem{Fayet:1977cr}
P.~Fayet and S.~Ferrara.
\newblock Supersymmetry.
\newblock {\em Phys. Rept.}, 32:249--334, 1977.

\bibitem{Barbieri:1988xf}
R.~Barbieri.
\newblock Looking beyond the standard model: The supersymmetric option.
\newblock {\em Riv. Nuovo Cim.}, 11N4:1--45, 1988.

\bibitem{Martin:1997ns}
S.~P. Martin.
\newblock {\em A supersymmetry primer}.
\newblock Perspectives on Supersymmetry. Advanced Series on Directions in High
  Energy Physics - Vol 18, World Scientific, 1998.

\bibitem{Bagger:1996ka}
J.~A. Bagger.
\newblock {\em Weak-scale supersymmetry: Theory and practice}.
\newblock Proceedings of the Boulder 1995 TASI Lectures (QCD161:T45:1995).

\bibitem{Haber:1985rc}
H.~E. Haber and G.~L. Kane.
\newblock The search for supersymmetry: Probing physics beyond the standard
  model.
\newblock {\em Phys. Rept.}, 117:75, 1985.

\bibitem{Jungman:1996df}
G.~Jungman, M.~Kamionkowski, and K.~Griest.
\newblock Supersymmetric dark matter.
\newblock {\em Phys. Rept.}, 267:195--373, 1996.

\bibitem{Nilles:1984ge}
H.~P. Nilles.
\newblock Supersymmetry, supergravity and particle physics.
\newblock {\em Phys. Rept.}, 110:1, 1984.

\bibitem{Chamseddine:1982jx}
A.~H. Chamseddine, R.~Arnowitt, and P.~Nath.
\newblock Locally supersymmetric grand unification.
\newblock {\em Phys. Rev. Lett.}, 49:970, 1982.

\bibitem{Barbieri:1982eh}
R.~Barbieri, S.~Ferrara, and C.~A. Savoy.
\newblock Gauge models with spontaneously broken local supersymmetry.
\newblock {\em Phys. Lett.}, B119:343, 1982.

\bibitem{Hall:1983iz}
L.~J. Hall, J.~Lykken, and S.~Weinberg.
\newblock Supergravity as the messenger of supersymmetry breaking.
\newblock {\em Phys. Rev.}, D27:2359--2378, 1983.

\bibitem{Gamberini:1990jw}
G.~Gamberini, G.~Ridolfi, and F.~Zwirner.
\newblock On radiative gauge symmetry breaking in the minimal supersymmetric
  model.
\newblock {\em Nucl. Phys.}, B331:331--349, 1990.

\bibitem{Feng:1999zg}
J.~L. Feng, K.~T. Matchev, and T.~Moroi.
\newblock Focus points and naturalness in supersymmetry.
\newblock {\em Phys. Rev.}, D61:075005, 2000.

\bibitem{Feng:2000gh}
J.~L. Feng, K.~T. Matchev, and F.~Wilczek.
\newblock Neutralino dark matter in focus point supersymmetry.
\newblock {\em Phys. Lett.}, B482:388--399, 2000.

\bibitem{Suspect}
A.~Djouadi, J.L. Kneur, and G.~Moultaka.
\newblock Suspect program,
  http://www.lpm.univ-montp2.fr:7082/~kneur/suspect.html.

\bibitem{Djouadi:2001yk}
A.~Djouadi, M.~Drees, and J.~L. Kneur.
\newblock Constraints on the minimal supergravity model and prospects for susy
  particle production at future linear e+ e- colliders.
\newblock {\em JHEP}, 08:055, 2001.

\bibitem{Castano:1994ri}
D.~J. Castano, E.~J. Piard, and Pierre Ramond.
\newblock Renormalization group study of the standard model and its extensions.
  2. the minimal supersymmetric standard model.
\newblock {\em Phys. Rev.}, D49:4882--4901, 1994.

\bibitem{Barger:1994gh}
V.~D. Barger, M.~S. Berger, and P.~Ohmann.
\newblock The supersymmetric particle spectrum.
\newblock {\em Phys. Rev.}, D49:4908--4930, 1994.

\bibitem{Pierce:1997zz}
Damien~M. Pierce, Jonathan~A. Bagger, Konstantin~T. Matchev, and Ren-jie Zhang.
\newblock Precision corrections in the minimal supersymmetric standard model.
\newblock {\em Nucl. Phys.}, B491:3--67, 1997.

\bibitem{Gondolo:2000ee}
P.~Gondolo, J.~Edsjo, L.~Bergstrom, P.~Ullio, and E.~A. Baltz.
\newblock Darksusy: A numerical package for dark matter calculations in the
  mssm, astro-ph/0012234, proceedings of york 2000, the identification of dark
  matter.
\newblock pages 318--323.

\bibitem{Darksusy}
P.~Gondolo, J.~Edsjö, L.~Bergström, P.~Ullio, and T.~Baltz.
\newblock Darksusy program, http://www.physto.se/~edsjo/darksusy/.

\bibitem{Nihei:2001qs}
T.~Nihei, L.~Roszkowski, and R.~Ruiz~de Austri.
\newblock Towards an accurate calculation of the neutralino relic density.
\newblock {\em JHEP}, 05:063, 2001.

\bibitem{Allanach:2001hm}
B.~C. Allanach.
\newblock Theoretical uncertainties in sparticle mass predictions.
  hep-ph/0110227, submitted to aps / dpf / dpb summer study on the future of
  particle physics (snowmass 2001), snowmass, colorado.
\newblock 30 Jun - 21 Jul 2001.

\bibitem{Allanach:2001kg}
B.~C. Allanach.
\newblock Softsusy: A c++ program for calculating supersymmetric spectra.
\newblock {\em Comput. Phys. Commun.}, 143:305--331, 2002.

\bibitem{PDG}
D.~E. Groom et~al.
\newblock Review of particle physics.
\newblock {\em Eur. Phys. J.}, C15:1--878, 2000.

\bibitem{Knecht:2001qf}
Marc Knecht and Andreas Nyffeler.
\newblock Hadronic light-by-light corrections to the muon g-2: The pion-pole
  contribution.
\newblock {\em Phys. Rev.}, D65:073034, 2002.

\bibitem{Czarnecki:2001pv}
Andrzej Czarnecki and William~J. Marciano.
\newblock The muon anomalous magnetic moment: A harbinger for 'new physics'.
\newblock {\em Phys. Rev.}, D64:013014, 2001.

\bibitem{Feng:2000zu}
J.~L. Feng, K.~T. Matchev, and F.~Wilczek.
\newblock Prospects for indirect detection of neutralino dark matter.
\newblock {\em Phys. Rev.}, D63:045024, 2001.

\bibitem{Bertolini:1991if}
S.~Bertolini, F.~Borzumati, A.~Masiero, and G.~Ridolfi.
\newblock Effects of supergravity induced electroweak breaking on rare b decays
  and mixings.
\newblock {\em Nucl. Phys.}, B353:591--649, 1991.

\bibitem{Ciuchini:1998xe}
M.~Ciuchini, G.~Degrassi, P.~Gambino, and G.~F. Giudice.
\newblock Next-to-leading qcd corrections to b --> x/s gamma: Standard model
  and two-higgs doublet model.
\newblock {\em Nucl. Phys.}, B527:21--43, 1998.

\bibitem{Barate:2000na}
R.~Barate et~al.
\newblock Search for the neutral higgs bosons of the standard model and the
  mssm in e+ e- collisions at s**(1/2) = 189-gev.
\newblock {\em Eur. Phys. J.}, C17:223--240, 2000.

\bibitem{Heister:2001kr}
A.~Heister et~al.
\newblock Final results of the searches for neutral higgs bosons in e+ e-
  collisions at s**(1/2) up to 209-gev.
\newblock {\em Phys. Lett.}, B526:191--205, 2002.

\bibitem{Boehm:1999bj}
C.~Boehm, A.~Djouadi, and M.~Drees.
\newblock Light scalar top quarks and supersymmetric dark matter.
\newblock {\em Phys. Rev.}, D62:035012, 2000.

\bibitem{Ellis:2001nx}
J.~R. Ellis, K.~A. Olive, and Y.~Santoso.
\newblock Calculations of neutralino stop coannihilation in the cmssm,
  cern-th-2001-339, umn-th-2032-01, tpi-minn-01-50, hep-ph/0112113.

\bibitem{Abusaidi:2000wg}
R.~Abusaidi et~al.
\newblock Exclusion limits on the wimp nucleon cross-section from the cryogenic
  dark matter search.
\newblock {\em Phys.Rev.Lett.}, 84:5699--5703, 2000.

\bibitem{Benoit:2002hf}
A.~Benoit et~al.
\newblock Improved exclusion limits from the edelweiss wimp search,
  astro-ph/0206271.

\bibitem{Edelweiss}
G.~Chardin.
\newblock Edelweiss dark matter search, talk given at the school and workshop
  on neutrino particle astrophysics, les houches 21 jan -1st feb 2002.

\bibitem{Zeplin}
N.~Spooner.
\newblock New limits and progress from the boulby dark matter programme, talk
  given at the school and workshop on neutrino particle astrophysics, les
  houches 21 jan -1rst feb.
\newblock 2002.

\bibitem{Mayet:2002ke}
F.~Mayet, D.~Santos, Yu.~M. Bunkov, E.~Collin, and H.~Godfrin.
\newblock Search for supersymmetric dark matter with superfluid he-3 (mache3).
\newblock {\em Phys. Lett.}, B538:257, 2002.

\bibitem{Suvorova:1999my}
Olga~V. Suvorova.
\newblock Status and perspectives of indirect search for dark matter, published
  in tegernsee 1999, beyond the desert 1999.
\newblock pages 853--867.

\bibitem{Macro}
T.~Montaruli.
\newblock Search for wimps using upward-going muons in macro, proceeedings of
  the 26th icrc in salt lake city, hep-ex/9905021.
\newblock pages 277--280, 17-25 Aug 1999.

\bibitem{SuperK}
A.~Habig.
\newblock Discriminating between nu/mu <--> nu/tau and nu/mu <--> nu(sterile)
  in atmospheric nu/mu oscillations with the super-kamiokande detector,
  proceedings of the 27th icrc, hamburg, germany, hep-ex/0106024.
\newblock 7-15 Aug 2001.

\bibitem{AntarLee}
L.~Thompson.
\newblock Dark matter prospects with the antares neutrino telescope, talk given
  at the conference dark 2002, cape town, south africa 4-9 feb.
\newblock 2002.

\bibitem{Ice3Edsjo}
J.~Edsjo.
\newblock Swedish astroparticle physics, talk given at the conference
  'partikeldagarna', uppsala, sweden, march 6.
\newblock 2001.

\bibitem{Ellis:1999mm}
J.~R. Ellis, T.~Falk, K.~A. Olive, and M.~Srednicki.
\newblock Calculations of neutralino stau coannihilation channels and the
  cosmologically relevant region of mssm parameter space.
\newblock {\em Astropart. Phys.}, 13:181--213, 2000.

\bibitem{mynext}
V.~Bertin, E.~Nezri, and J.~Orloff.
\newblock in preparation.

\bibitem{Bergstrom:1998xh}
Lars Bergstrom, Joakim Edsjo, and Paolo Gondolo.
\newblock Indirect detection of dark matter in km-size neutrino telescopes.
\newblock {\em Phys. Rev.}, D58:103519, 1998.

\bibitem{Ellis:2000ds}
John~R. Ellis, Andrew Ferstl, and Keith~A. Olive.
\newblock Re-evaluation of the elastic scattering of supersymmetric dark
  matter.
\newblock {\em Phys. Lett.}, B481:304--314, 2000.

\bibitem{Ellis:2001qm}
John~R. Ellis, Andrew Ferstl, and Keith~A. Olive.
\newblock Constraints from accelerator experiments on the elastic scattering of
  cmssm dark matter.
\newblock {\em Phys. Lett.}, B532:318--328, 2002.

\bibitem{Barger:2001ur}
Vernon~D. Barger, Francis Halzen, Dan Hooper, and Chung Kao.
\newblock Indirect search for neutralino dark matter with high energy
  neutrinos.
\newblock {\em Phys. Rev.}, D65:075022, 2002.

\bibitem{Roszkowski:2001sb}
L.~Roszkowski, R.~Ruiz~de Austri, and T.~Nihei.
\newblock New cosmological and experimental constraints on the cmssm.
\newblock {\em JHEP}, 08:024, 2001.

\bibitem{Baer:1999sp}
Howard Baer, Frank~E. Paige, Serban~D. Protopopescu, and Xerxes Tata.
\newblock Isajet 7.48: A monte carlo event generator for p p, anti-p p, and e+
  e- reactions, bnl-het-99-43, fsu-hep-991218, uh-511-952-00, hep-ph/0001086.
\newblock 1999.

\end{thebibliography}
\bibliographystyle{unsrt}

\end{document}